\documentclass[twocolumn,showpacs,preprintnumbers,amsmath,amssymb,superscriptaddress]{revtex4}

\usepackage{graphicx}
\usepackage{dcolumn}
\usepackage{bm}
\usepackage{color}
\usepackage{amsmath, amssymb}
\usepackage{type1cm}

\begin{document}

\preprint{APS}

\title{Response of the Higgs amplitude mode \\
of superfluid Bose gases in a three dimensional optical lattice}

\author{Kazuma Nagao}
\email{kazuma.nagao@yukawa.kyoto-u.ac.jp}
\affiliation{%
Yukawa Institute for Theoretical Physics, Kyoto University, Kitashirakawa Oiwakecho, Sakyo-ku, Kyoto 606-8502, Japan
}%
\author{Yoshiro Takahashi}%
\affiliation{%
Department of Physics, Kyoto University, Kitashirakawa Oiwakecho, Sakyo-ku, Kyoto 606-8502, Japan
}%
\author{Ippei Danshita}%
\affiliation{%
Yukawa Institute for Theoretical Physics, Kyoto University, Kitashirakawa Oiwakecho, Sakyo-ku, Kyoto 606-8502, Japan
}%

\date{\today}

\begin{abstract}
We study the Higgs mode of superfluid Bose gases in a three dimensional optical lattice, which emerges near the quantum phase transition to the Mott insulator at commensurate fillings. Specifically, we consider responses of the Higgs mode to temporal modulations of the onsite interaction and the hopping energy. In order to calculate the response functions including the effects of quantum and thermal fluctuations, we map the Bose-Hubbard model onto an effective pseudospin-one model and use a perturbative expansion based on the imaginary-time Green's function theory. We also include the effects of an inhomogeneous trapping potential by means of a local density approximation. We find that the response function for the hopping modulation is equal to that for the interaction modulation within our approximation. At the unit filling rate and in the absence of a trapping potential, we show that the Higgs mode can exist as a sharp resonance peak in the dynamical susceptibilities at typical temperatures. However, the resonance peak is significantly broadened due to the trapping potential when the modulations are applied globally to the entire system. We suggest that the Higgs mode can be detected as a sharp resonance peak by partial modulations around the trap center.

\end{abstract}

\pacs{Valid PACS appear here}
\maketitle

\section{\label{sec:level1}Introduction}

The Higgs amplitude mode is one of the universal quasi-particle excitations of thermodynamic phases with a particle-hole symmetry and spontaneous breaking of a continuous symmetry \cite{Volovik_2014, Pekker_2015}. In an intuitive picture, this mode corresponds to a massive fluctuation mode of the amplitude of the order parameter. Moreover, the Higgs mode is an analog of the Higgs boson in particle physics \cite{Higgs_1964}. The ubiquity of the Higgs mode in quantum many-body systems has attracted particular attention from many experimental research fields of condensed matter and ultracold gases \cite{Pekker_2015}. The examples known so far include superconductors NbSe$_2$ \cite{Sooryakumar_1980, Sooryakumar_1981, Littlewood_1981, Littlewood_1982,Measson_2014} and Nb$_{1-x}$Ti$_{x}$N \cite{Matsunaga_2013,Matsunaga_2014,Sherman_2015,Matsunaga_2017}, quantum antiferromagnets TlCuCl$_3$ \cite{Ruegg_2008, Merchant_2014} and KCuCl$_3$ \cite{Kuroe_2014}, charge density wave materials K$_{0.3}$MoO$_3$ \cite{Demsar_1999,Schaefer_2014} and TbTe$_3$ \cite{Yusupov_2010, Mertelj_2013}, superfluid ${^3}$He B-phase \cite{Avenel_1980, Collett_2013}, and superfluid Bose gases in optical lattices \cite{Bissbort_2011,Endres_2012}.

In the case of Bose gases in optical lattices, the Higgs mode is expected to appear in the superfluid phase at commensurate filling rates and near a critical value of the lattice depth at which the superfluid to Mott-insulator transition occurs \cite{Bissbort_2011, Endres_2012, Pekker_2015}. The Max-Planck group has experimentally explored the Higgs mode of Bose gases in a two-dimensional (2D) optical lattice by utilizing the lattice-amplitude modulation and the quantum-gas microscope technique \cite{Endres_2012}. They observed the energy gap of the Higgs mode by measuring a response of the system to the temporal modulation of the lattice amplitude as a function of the frequency. Although the measured energy gap agrees with the Higgs gap computed theoretically, the response versus the frequency exhibits a broad continuum above the gap energy rather than a sharp peak. In this sense, it remains as an open issue whether the Higgs mode in the optical-lattice system can exist as a well-defined quasiparticle.

The Max-Planck experiment \cite{Endres_2012} has stimulated detailed studies on addressing the issue, in particular, theoretical calculations in the 2D relativistic $O(N)$ scalar model \cite{Podolsky_2011, Podolsky_2012, Gazit_2013, Rancon_2014, Katan_2015, Rose_2015} or the 2D Bose-Hubbard model \cite{Pollet_2012, Chen_2013, Liu_2015}. The quantum Monte-Carlo simulations \cite{Pollet_2012, Liu_2015} in the presence of a trapping potential and at finite temperatures have shown that the linear response function to the lattice amplitude modulation exhibits no resonance peak at the Higgs energy gap. This result implies that the Higgs mode becomes unstable due to the combined effects of the quantum and thermal fluctuations and the spatial inhomogeneity of the trapping potential. Thus, it may be difficult to observe the Higgs mode as a well-defined quasiparticle excitation in the 2D optical-lattice systems.

In superfluid Bose gases in a three-dimensional (3D) optical lattice, in contrast to the 2D systems, we expect the existence of more stable Higgs modes because of the general fact that the long-range order of the systems becomes more robust against fluctuations as the spatial dimension increases. One of quantities characterizing the stability of the Higgs mode is its damping rate \cite{Altman_2002, Nagao_2016}. Altman and Auerbach have calculated the damping rate at zero temperature by means of the mapping of the Bose-Hubbard model at large filling rates to the effective pseudospin-1 model \cite{Altman_2002}. Thereafter, the current authors have generalized their zero temperature analysis to the finite-temperature case by applying the finite-temperature Green's function theory for the effective model \cite{Nagao_2016}. The latter result revealed that the Higgs modes are underdamped even at typical experimental temperatures. 

While the damping rate is a useful quantity for characterizing theoretically the stability of the Higgs mode, it is rather difficult to measure directly the damping rate in cold-atom experiments. In typical experiments, such as the Max-Planck experiment \cite{Endres_2012}, some response functions to a temporal modulation of an external field have been measured. In Ref. \cite{Huber_2007}, a response function of the 3D Bose-Hubbard model in the presence of a parabolic trapping potential to the lattice-amplitude modulation has been theoretically analyzed by means of the mapping to the modified effective pseudospin-1 model at lower filling rates, linear response theory, and local density approximation. It has been shown that within an approximation ignoring any fluctuation effects there exists a sharp resonance peak at the Higgs energy gap in a response function of the superfluid phase. There the broadening of the peak width stems only from the spatial inhomogeneity. As a next step toward understanding the detectability of the Higgs mode in the 3D systems, we should evaluate quantum and thermal fluctuation effects on the response functions.

In this work, we study effects of the lowest order correction with respect to the fluctuations to some response functions of the 3D Bose-Hubbard model in the presence of a parabolic trapping potential. In order to take into account fluctuation effects at lower filling rates, we apply the field theoretical approach developed in our previous work at a large filling limit \cite{Nagao_2016} for the modified effective pseudospin-1 model \cite{Huber_2007}. In particular, we concentrate on the case where the mean density at the trap center is equal to unity. In addition, we include the trapping-potential effect by using the local density approximation. At the unit filling rate and in the absence of the trapping potential, the dynamical susceptibilities show that the Higgs mode can exist as a sharp resonance peak at typical temperatures. In contrast, when we take into account the trapping potential and modulate the system globally, the resonance peak turns to be broadened significantly due to the inhomogeneity. To obtain a sharp peak in the presence of the trapping potential, we discuss partial modulations around the trap center, which have been analyzed also in the previous work \cite{Liu_2015} for 2D systems. We suggest that the Higgs mode is detectable as a sharp resonance peak in the presence of the trapping potential when we modulate the system with a modulation radius $R_{\rm mod}<0.5R_{\rm TF}$, where $R_{\rm TF}$ is the Thomas--Fermi radius of the trapped condensate.

The organization of this paper is as follows. In Sec. \ref{Sec: Model}, we introduce the tight-binding Bose-Hubbard model and formulate a linear response theory. In Sec. \ref{Sec: Methods}, we explain a method approximately describing the low-energy properties of the superfluid near the Mott-insulator transition on the basis of the mapping of the Bose-Hubbard model into the effective pseudospin-one model. In Sec. \ref{Sec: LRA}, we discuss how to compute the response functions within the method developed in Sec. \ref{Sec: Methods}. In particular, we calculate the response functions by using the framework of the imaginary-time Green's function theory. In Sec. \ref{Sec: Results}, we show the frequency dependence of the imaginary part of the susceptibilities both in the absence and in the presence of the trapping potential. We discuss whether the Higgs mode can exist as a well-defined sharp resonance peak in the quantities. In addition, we also discuss the finite-temperature effects on the results at zero temperature. In Sec. \ref{Sec: conclusions}, we summarize the results and describe our outlook. Throughout this paper, we set the reduced Planck constant $\hbar$, the lattice spacing $d_{\rm lat}$, and the Boltzmann constant $k_{\rm B}$ as units: $\hbar=d_{\rm lat} = k_{\rm B} = 1$.

\section{Bose-Hubbard model}\label{Sec: Model}

In this paper, we investigate the collective fluctuation modes of superfluid Bose gases in a cubic optical lattice in the presence of a parabolic trapping potential. When the lattice depth is sufficiently deep, the system can be described by the tight-binding Bose-Hubbard model \cite{Fisher_1989, Jaksch_1998}
\begin{align}
{\cal H}_{\rm BH}=-J\sum_{\langle ij \rangle}a_{i}^{\dagger}a_{j}+\frac{U}{2}\sum_{i}(n_i-n_0)^2 - \sum_{i}\mu_{i}(n_i-n_0), \nonumber
\end{align}
where $a_{i}$ and $a^\dagger_{i}$ are boson annihilation and creation operators at site $i$ of the cubic lattice, $\langle ij \rangle$ denotes a summation over all possible nearest-neighbor pairs of the sites, $n_i=a_i^\dagger a_i$ is the density operator at site $i$, and $n_0$ is a nonzero and positive integer. This notation for the Bose-Hubbard model is suitable for our approximation around the $n_0$th Mott-insulator region (see Sec. \ref{Sec: Effective}). The parameters $J$, $U$, and $\mu_i$ are the hopping strength, onsite-interaction strength, and local chemical potential. When the trapping potential is $V_{\rm trap}(r)$, and the chemical potential at the center of the potential is $\mu$, the local chemical potential is given by $\mu_i = \mu - V_{\rm trap}(r)$. Now $r$ is the radial distance measured from the center. 

In this section and the subsequent sections from Sec. \ref{Sec: Methods} to Sec. \ref{Sec: RFs_unif}, we confine ourselves to the spatially homogeneous case for simplicity; i.e., we concentrate on studying the bulk properties. Effects of the trapping potential will be discussed within the local density approximation in Sec. \ref{Sec: LDA}.

At an integer (or commensurate) filling rate, the Bose-Hubbard model has two different ground states, i.e. superfluid and Mott-insulator states \cite{Fisher_1989, Oosten_2001,Sansone_2007}. The phase boundary corresponds to a critical value of the dimensionless ratio $Jz/U$, where $z=2d=6$ is the coordination number. For $d = 3$, the superfluid to Mott-insulator transition is of second order \cite{Fisher_1989, Oosten_2001, Sansone_2007}. Its universality class belongs to that of the $(d+1)$-dimensional classical $XY$ model \cite{Fisher_1989}. 

In the vicinity of the critical point with an integer filling rate, the dynamical critical exponent becomes $z_{\rm dyn}=1$ \cite{Fisher_1989}. There, the corresponding effective action has the form of the relativistic $O(2)$ field theory \cite{Sachdev_2011, Nakayama_2015}. Because of the second-order time derivative term, the phase and amplitude fluctuations are no longer canonical conjugate with each other. Thus, the amplitude fluctuation and its conjugate momentum (not the phase fluctuation) form one collective mode, i.e., the gapped Higgs amplitude mode, which is independent of the phase fluctuation. In a similar way, the phase fluctuation and its conjugate momentum also form the gapless Nambu--Goldstone (NG) phase mode independently. This is in contrast to the nonrelativistic Gross-Pitaevskii case, where the phase fluctuation is canonical conjugate with the amplitude one. In this case, these degrees of freedom form only one collective mode, namely the gapless Bogoliubov mode. For more detailed discussion, see Ref. \cite{Altman_2015}.

\section{External perturbations}\label{Sec: external perturbations}

In this section, we discuss external perturbations that are time dependent and coupled with the Higgs mode in the optical-lattice system. We formulate the responses to the perturbations within the linear response theory.

The basic idea for exciting the Higgs mode is to modulate the condensate density $|\Psi|^2$ with a small amplitude of vibration \cite{Pekker_2015}. For the Bose-Hubbard model, this can be performed by modulating a dimensionless ratio $J/U$, which determines the order-parameter amplitude of a ground state. A typical method utilized in experiments to modulate it is the optical-lattice amplitude modulation technique \cite{Stoferle_2004, Endres_2012}, which leads to a modulation of the hopping strength $J$ (its detailed discussions can be found in some literatures \cite{Pekker_2015, Endres_thesis, Liu_2015}). The experiment of Ref. \cite{Endres_2012} has achieved periodic modulations of the lattice depth with a sufficiently small vibrational amplitude (3 \% of the initial depth) to the extent that the resulting response is in a linear response region. 

\subsection{Modulations of the kinetic energy}\label{Sec: modkin}

The response to the $J$ modulation can be formulated by the linear response theory as follows: Let us assume that the system is in the thermal equilibrium state with the inverse temperature $\beta = T^{-1}$ at $t \rightarrow -\infty$. When we add a small and periodic modulation to the hopping strength $J$ slowly such that $J \rightarrow [1+\Delta_J(t)]J $ where $\Delta_J(t) = \delta_J {\rm cos}(\omega t)$ and $\delta_J $ is sufficiently small, then the Hamiltonian describing the weak perturbation reads
\begin{align}
{\cal H}_{\rm BH} \rightarrow {\cal H}_{\rm BH} + \Delta_{J}(t) K, \nonumber
\end{align}
where $K \equiv -J \sum_{\langle ij \rangle}a^{\dagger}_{i}a_{j}$ is the kinematic energy. The second term on the righthand side denotes the weak perturbation term. The instantaneous change of the total energy to the small and periodic modulations is proportional to the instantaneous quantum mechanical average of the kinetic energy \cite{Pekker_2015, Endres_thesis}. Therefore, the response of the system to the modulations is characterized only by the response of the kinetic energy and described by the $K$-to-$K$ response function \cite{Pollet_2012, Liu_2015}
\begin{align}
D^{\rm R}_{KK}(t-t')&=-i\Theta(t-t') \left\langle \left[K(t),K(t')\right] \right\rangle_{\rm eq}, \label{eq: response_kk}
\end{align}
where $\Theta(t)$ is the step function, which outputs 1 for $t>0$ and 0 for $t<0$. Here, $K(t) = e^{i{\cal H}_{\rm BH}t}Ke^{-i{\cal H}_{\rm BH}t}$. The bracket $\langle \cdots \rangle_{\rm eq}$ means the normalized ensemble average of the thermal equilibrium state at $t \rightarrow -\infty$: $\langle \cdots \rangle_{\rm eq} \equiv {\rm Tr}(e^{-{\beta \cal H}_{\rm BH}}\cdots )/{\rm Tr}\;e^{-{ \beta \cal H}_{\rm BH}}$. The imaginary part of the dynamical susceptibility
\begin{align}
\chi_{KK}(\omega) = \int^{\infty}_{-\infty} D^{\rm R}_{KK}(t)e^{i \omega t}dt,\label{eq: susceptibility_kk}
\end{align}
is the spectral function $S_{KK}(\omega)=-{\rm Im}\left[\chi_{KK}(\omega)\right]$, which is proportional to the external energy absorbed by the system for a finite-time period of the modulation \cite{Pekker_2015, Endres_2012}. The response function or its susceptibility characterizes the resonance of the Higgs mode in experimental systems \cite{Endres_2012,Pollet_2012}. The Max-Planck experiment \cite{Endres_2012} has obtained $S_{KK}(\omega)$ at low frequencies by measuring the temperature increase of the system after the lattice-amplitude modulation with a fixed modulation time.

\subsection{Modulations of the onsite-interaction energy}\label{Sec: modons}

In Sec. \ref{Sec: modkin}, we briefly reviewed the conventional $J$ modulations and consequent response. On the other hand, one can also modulate the onsite interaction $U$ to oscillate $J/U$. To our knowledge, this kind of modulation has not been discussed thus far as a probe of the Higgs mode. In this section we explain what types of response function characterize the response to the $U$ modulations and how one can realize that modulation in experiments with high controllability. Moreover, we will show in detail  the relation between the response function and energy absorbed by the system for a period of the $U$ modulation in Appendix \ref{App: onsite}.

Let us consider a linear response problem to the $U$ modulation in a similar way to the $J$ modulation. When we turn on a small and periodic modulation $U \rightarrow [1 + \Delta_U(t)]U$ where $\Delta_U(t) = \delta_U {\rm cos}(\omega t)$ and $\delta_U $ is sufficiently small, then the Hamiltonian becomes
\begin{align}
{\cal H}_{\rm BH} \rightarrow {\cal H}_{\rm BH} + \Delta_{U}(t) O, \nonumber
\end{align}
where $O \equiv \frac{U}{2}\sum_i (n-n_0)^2$ is the onsite-interaction energy. In a manner similar to the $J$ modulations, the instantaneous change rate of the total energy is proportional to the quantum mechanical average of the onsite energy (for details, see Appendix \ref{App: onsite}). Thus, within the linear response theory, the consequent response can be described by the $O$-to-$O$ response function
\begin{align}
D^{\rm R}_{OO}(t-t')&=-i \Theta(t-t') \left\langle \left[O(t),O(t')\right] \right\rangle_{\rm eq}, \label{eq: response_oo}
\end{align}
where $O(t) = e^{i{\cal H}_{\rm BH}t}Oe^{-i{\cal H}_{\rm BH}t}$. The imaginary part of the dynamical susceptibility
\begin{align}
\chi_{OO}(\omega) = \int^{\infty}_{-\infty} D^{\rm R}_{OO}(t)e^{i\omega t}dt, \label{eq: susceptibility_oo}
\end{align}
is the spectral function $S_{OO}(\omega)=-{\rm Im}\left[\chi_{OO}(\omega)\right]$, which is proportional to the external energy absorbed by the system for a finite-time period of the modulation (see Appendix \ref{App: onsite}). We expect that this response function or its susceptibility also characterizes the resonance of the Higgs mode. The difference with $\chi_{KK}(\omega)$ will be discussed in Sec. \ref{Sec: LDA}.

Recent experimental developments in the fields of ultracold gases enable one to control the onsite interaction by using highly controlled optical techniques, such as {\it the optical Feshbach resonance} \cite{Fedichev_1996, Theis_2004, Yamazaki_2010} and {\it the optically induced Feshbach resonance} \cite{Bauer_2009, Clark_2015}. In contrast to the conventional magnetic Feshbach resonance, these techniques allow for fast temporal modulation of $U$ with a frequency on the order of 1 to 10 kHz, which is supposed to be comparable to a typical resonance frequency of the Higgs mode.

\section{Methods}\label{Sec: Methods}

In order to analyze the Higgs mode, we use the mapping of the Bose-Hubbard model onto an effective pseudospin-1 model \cite{Altman_2002,Huber_2007} and field theoretical method based on the imaginary-time Green's function. This section is devoted to explaining how to describe collective modes of the superfluid phase beginning with the effective pseudospin-one model. The application of the field theoretical method for computing the response functions of the system will be discussed in Sec. \ref{Sec: LRA}.

\subsection{Effective pseudospin-one model near the Mott-insulator transition}\label{Sec: Effective}

Let us discuss an effective description of the superfluid state with a commensurate filling rate $n_0$. In the vicinity of the Mott-insulator transition, the local fluctuations of $n_i$ from the mean density $\langle n_i \rangle = n_0$ are sufficiently suppressed. Therefore, low-energy properties of the system can be described by an effective model 
\begin{align}
{\cal H}^{n_0}_{\rm eff}={\cal P}_{n_0} {\cal H}_{\rm BH} {\cal P}_{n_0}^{-1}, \label{eq: def_eff} 
\end{align}
where ${\cal P}_{n_0}$ is a projection operator eliminating high-energy Fock states $|n_0+\alpha\rangle$ for $|\alpha|>1$ from the complete Hilbert space. The remaining states, which describe the low-energy phenomena effectively, can be represented by three Schwinger bosons \cite{Altman_2002, Huber_2007}
\begin{align}
|n_0+\alpha \rangle_i \equiv t_{\alpha,i}^{\dagger}|{\rm vac}\rangle,\;\;{\rm for}\;\;\alpha=-1,0,1, \nonumber 
\end{align}
where $|{\rm vac}\rangle$ is the vacuum of new bosons. The commutation relations are $[t_{\alpha,i},t^\dagger_{\alpha',j}]=\delta_{\alpha,\alpha'}\delta_{i,j}$ and $[t_{\alpha,i},t_{\alpha',j}]=[t^\dagger_{\alpha,i},t^\dagger_{\alpha',j}]=0$. In order to eliminate the unphysical states such as $t^\dagger_{1,i}t^\dagger_{0,i}|{\rm vac}\rangle$, we assume that these operators obey a constraint 
\begin{align}
\sum_{\alpha=-1}^{1} t^{\dagger}_{\alpha,i} t_{\alpha,i}= {\hat 1}, \label{eq: constraint1}
\end{align}
where ${\hat 1}$ on the right-hand side is the identity operator in the reduced Hilbert subspace.

For sufficiently large filling rates ($n_0 \gg 1$), the effective model becomes a simple pseudospin-1 model \cite{Altman_2002}
\begin{align}
{\cal H}^{n_0\gg1}_{\rm eff}=-\frac{Jn_0}{2}\sum_{\langle ij \rangle}S_i^{+}S_j^- +\frac{U}{2}\sum_i (S_i^z)^2 - B \sum_i S_i^z, \label{eq: high filling efm}
\end{align}
where $B = \mu$ is the uniform {\it magnetic field} coupling with the $z$-component of the pseudospins. The pseudospin-1 operators are defined by
\begin{align}
S_i^+ &= \sqrt{2}(t_{1,i}^{\dagger}t_{0,i}+t_{0,i}^{\dagger}t_{-1,i}), \nonumber \\
S_i^- &= \sqrt{2}(t_{0,i}^{\dagger}t_{1,i}+t_{-1,i}^{\dagger}t_{0,i}), \nonumber \\
S_i^z &= t_{1,i}^{\dagger}t_{1,i}-t_{-1,i}^{\dagger}t_{-1,i}, \nonumber 
\end{align}
and satisfy the SU(2) commutation relations $[S^+_i,S^-_j]=2S^z_i \delta_{i,j}$ and $[S^z_i,S^{\pm}_j] = \pm S^{\pm}_i\delta_{i,j}$. Note that the $XY$ spin exchange, on-site single-ion anisotropy, and magnetic coupling terms in the effective model correspond to the hopping, onsite-interaction, and chemical potential terms in the Bose-Hubbard model, respectively. The effective model (\ref{eq: high filling efm}) has a particle-hole symmetry at a commensurate filling rate corresponding to $B = \mu=0$. We will find later that this particle-hole symmetry forbids interactions associated with an odd number of the NG mode. For details, see Sec. \ref{Sec: Bogoliubov}.

The large-filling model (\ref{eq: high filling efm}) is not adequate for quantitatively describing typical experimental situations with lower filling rates. In fact, the mean filling rate at the center of the trap in the Max-Planck experiment \cite{Endres_2012} was tuned to be unity. For lower commensurate filling rates ($n_0 \sim 1$), we need to modify the spin exchange term \cite{Huber_2007} such that
\begin{align}
{\cal H}^{n_0}_{\rm eff}= & -\frac{Jn_0}{2}\sum_{\langle ij \rangle}(1+\delta \nu S_i^z)S_i^+ S_j^-(1+\delta \nu S_j^z)  \nonumber \\
&+\frac{U}{2}\sum_i (S_i^z)^2 - B \sum_i S_i^z, \label{eq: low filling efm}
\end{align}
where $\delta \nu = \sqrt{1+1/n_0} - 1$. The modified model (\ref{eq: low filling efm}) has no longer the particle-hole symmetry even at a commensurate filling rate. Nevertheless, the Higgs mode can exist as an independent collective mode even at low filling rates as long as the system is near the transition to the Mott insulating phase. This happens because an effective particle-hole symmetry emerges in such a region.

The absence of the particle-hole symmetry makes it complicated to compute the fluctuation correction of the response functions because no constraint forbids interactions associated with an odd number of the NG modes. For details, see Sec. \ref{Sec: Bogoliubov}. Note that $\delta \nu$ measures the deviation from the particle-hole symmetric point. Obviously, if $\delta \nu \rightarrow 0$, the effective model then approaches the particle-hole symmetric model (\ref{eq: high filling efm}).

In this paper, in order to obtain the response functions corresponding to typical experiments, we mainly use the latter model at the unit filling rate. The former model will be used for calculating the large-filling response functions in the absence of the trapping potential and at zero temperature. In Sec. \ref{Sec: RFs_unif}, we compare two limiting results of the unit filling rate and a large filling rate in such a situation.

\subsection{Mean-field ground state in the truncated Hilbert subspace}\label{Sec: MFGS}

In this subsection, we make an ansatz of the ground state wave function of the effective pseudospin-1 model, which is essentially equivalent to a mean-field approximation of the ground state in the truncated Hilbert subspace, according to Refs. \cite{Altman_2002, Huber_2007}. 

We define a variational wave function of the ground state as
\begin{align}
|\Omega (\theta,\eta,\varphi,\chi) \rangle = \prod_i \left\{ {\rm cos}\left(\frac{\theta}{2}\right)t_{0,i}^{\dagger}  + e^{{\rm i}\eta}{\rm sin}\left(\frac{\theta}{2}\right) \right. \nonumber \\
\left. \times \left[ e^{{\rm i}\varphi}{\rm sin}\left(\frac{\chi}{2}\right) t_{1,i}^{\dagger}+e^{-{\rm i}\varphi}{\rm cos}\left(\frac{\chi}{2}\right)t_{-1,i}^{\dagger} \right] \right\}| {\rm vac}\rangle, \label{eq: Gutzwiller}
\end{align}
where $\theta \in [0,\pi],\;\eta \in [-\pi/2,\pi/2],\;\varphi \in [0,2\pi],\;{\rm and}\;\chi \in [0,\pi]$ are the variational parameters. Note that this wave function at $\theta = 0$ describes the Mott-insulating state of $n_0$ filling factor with no fluctuation, i.e., $\prod_{i}t^\dagger_{0,i}|{\rm vac}\rangle$. In the superfluid phase, $\theta \neq 0$ mixes the mean filling state $t^\dagger_{0,i}$ with the particle and hole fluctuations $t^\dagger_{1,i}$ and $t^\dagger_{-1,i}$. Hence, it plays a role of the order parameter strength.

In the superfluid phase ($\theta \neq 0$), the variational parameters are determined from minimizing the mean energy density $E^{\rm MF} = \langle \Omega | {\cal H}^{n_0}_{\rm eff} |\Omega \rangle / N$ with respect to the variational parameters. Here, $N$ is the total number of the lattice point. A specific representation of $E^{\rm MF}$ for the ground state is shown in Appendix \ref{App: variational}.

From the Ginzburg--Landau expansion of $E^{\rm MF}$ of the ground state with respect to the order parameter $\Psi = \langle \Omega | a_i | \Omega \rangle$, we can determine the phase-boundary of the superfluid to insulator transition \cite{Huber_2007}. Now we introduce a dimensionless parameter $u=U/(4Jn_0z)$ measuring the distance from the critical point at the commensurate filling rate. The critical value of the superfluid to insulator transition within the mean-field approximation \cite{Huber_2007} is
\begin{align}
u_c = \frac{1}{4n_0}(\sqrt{n_0 + 1} + \sqrt{n_0})^2. \label{eq:critical value}
\end{align}
At $n_0 \rightarrow \infty$, the critical value $u_c$ approaches $1$. At the unit filling rate $n_0 = 1$, $u_c = (\sqrt{2}+1)^2/4 \approx 1.457$. Note that the same result can be obtained from the site-decoupling mean-field approximation of the Bose-Hubbard model \cite{Oosten_2001}. The exact critical value at the unit filling rate has been numerically computed as $u_c=1.22(2)$ by the quantum Monte-Carlo method of the 3D Bose-Hubbard model in Ref.~\cite{Sansone_2007}. In this paper, we mainly use the mean-field result of Eq.~(\ref{eq:critical value}) to be consistent with our analysis on the mean-field ground state.

\subsection{Fluctuations from the mean-field ground state}\label{Sec: Excitations}

In Sec.~\ref{Sec: MFGS} we have discussed the ground state properties of the effective pseudospin-1 model within the mean-field approximation. In this subsection, we turn to consider fluctuations arising on the mean-field superfluid state.

Based on the variational ansatz of the ground state wave function, we can formulate the collective excitations on the superfluid phase as fluctuations around the mean-field state \cite{Altman_2002, Huber_2007}. Let us introduce the creation operators of the mean-field ground state of the superfluid $|\Omega \rangle \equiv \prod_i b_{0,i}^\dagger|{\rm vac}\rangle$ and define a canonical transformation
\begin{align}
b_{0,i}^{\dagger}&= c_1 t_{0,i}^{\dagger} + s_1 \left[ s_2 t_{1,i}^{\dagger}  + c_2 t_{-1,i}^{\dagger}  \right], \nonumber \\
b_{1,i}^{\dagger}&= s_1 t_{0,i}^{\dagger} - c_1  \left[ s_2 t_{1,i}^{\dagger}  + c_2 t_{-1,i}^{\dagger}  \right],  \label{eq: canonical} \\ 
b_{2,i}^{\dagger}&= c_2 t_{1,i}^{\dagger} - s_2 t_{-1,i}^{\dagger}, \nonumber 
\end{align}
where the coefficients are $s_1 = {\rm sin}(\theta_{\rm mf}/2)$, $c_1 = {\rm cos}(\theta_{\rm mf}/2)$, $s_2 = {\rm sin}(\chi(\theta_{\rm mf})/2)$, and $c_2 = {\rm cos}(\chi(\theta_{\rm mf})/2)$. $\theta_{\rm mf}$ denotes the value of the variational parameter $\theta$ for the ground state. $b_{1,i}^{\dagger}$ describes the amplitude fluctuation of the order parameter on the ground state while $b_{2,i}^{\dagger}$ describes the phase fluctuation. These new operators fulfill the same commutation relations as the old operators $t_{\alpha,i}$. In addition, the transformation retains the constraint (\ref{eq: constraint1}) so that
\begin{align}
\sum_{m = 0}^{2} b^{\dagger}_{m,i} b_{m,i}= {\hat 1}. \label{eq: constraint2}
\end{align}

Substituting the canonical transformation (\ref{eq: canonical}) into the effective model (\ref{eq: low filling efm}), we obtain the Hamiltonian describing the collective fluctuations around the mean-field ground state. The resulting Hamiltonian  consists of five successive parts
\begin{align}
{\cal H}_{\rm eff} = {\cal H}^{(0)}_{\rm eff} + {\cal H}^{(1)}_{\rm eff} + {\cal H}^{(2)}_{\rm eff} + {\cal H}^{(3)}_{\rm eff} + {\cal H}^{(4)}_{\rm eff}, \label{eq: effective hamiltonian}
\end{align}
where each term contained in ${\cal H}^{(l)}_{\rm eff}$ ($l = 0,1,2,3,4$) has $l$ numbers of the fluctuation operator $b^\dagger_{m,i},b_{m,i}$ ($m=1,2$). The explicit form of ${\cal H}^{(l)}_{\rm eff}$ is given by 
\begin{widetext}
\begin{align}
{\cal H}^{(0)}_{\rm eff}= &\sum_{\langle ij \rangle}\frac{1}{z}A_0b^\dagger_{0,i} b_{0,i} b^\dagger_{0,j} b_{0,j} + \sum_i {\tilde A}_0b^\dagger_{0,i} b_{0,i},  \nonumber \\
{\cal H}^{(1)}_{\rm eff}= &\sum_{\langle ij \rangle} \frac{1}{z}A_{1} b^\dagger_{0,i} b_{0,i} b^\dagger_{1,j} b_{0,j}  +  \sum_{\langle ij \rangle} \frac{1}{z}B_{1} b^\dagger_{0,i} b_{0,i} b^\dagger_{2,j} b_{0,j} + \sum_i {\tilde A}_{1} b^\dagger_{1,i} b_{0,i} +  \sum_i {\tilde B}_{1} b^\dagger_{2,i} b_{0,i} + {\rm H.c.}, \nonumber  \\
{\cal H}^{(2)}_{\rm eff}=&\sum_{\langle ij \rangle}\frac{1}{2z}A_2 b^\dagger_{0,i} b_{0,i} b^\dagger_{1,j} b_{1,j} + \sum_{\langle ij \rangle}\frac{1}{z}B_2 b^\dagger_{0,i} b_{0,i} b^\dagger_{1,j}b_{2,j} + \sum_{\langle ij \rangle}\frac{1}{z}D_2 b^\dagger_{1,i} b_{0,i} b^\dagger_{1,j} b_{0,j} + \sum_{\langle ij \rangle}\frac{1}{2z}E_2 b^\dagger_{1,i} b_{0,i} b^\dagger_{0,j} b_{1,j} \nonumber \\
+&\sum_{\langle ij \rangle}\frac{1}{z}F_2 b^\dagger_{1,i} b_{0,i} b^\dagger_{0,j} b_{2,j} +  \sum_{\langle ij \rangle}\frac{1}{z}G_2 b^\dagger_{1,i} b_{0,i}  b^\dagger_{2,j} b_{0,j} + \sum_{\langle ij \rangle}\frac{1}{z}H_2 b^\dagger_{0,i} b_{2,i} b^\dagger_{0,j} b_{2,j} + \sum_{\langle ij \rangle}\frac{1}{2z}I_2 b^\dagger_{0,i} b_{2,i} b^\dagger_{2,j} b_{0,j} \nonumber \\
+&\sum_i \frac{1}{2}{\tilde A}_2 b^\dagger_{1,i} b_{1,i} + \sum_i {\tilde B}_2 b^\dagger_{1,i} b_{2,i} +  \sum_i \frac{1}{2}{\tilde C}_2b^\dagger_{2,i} b_{2,i} + {\rm H.c.}, \nonumber \\
{\cal H}^{(3)}_{\rm eff}= &\sum_{\langle ij \rangle}\frac{1}{z}A_3 b^\dagger_{1,i} b_{1,i} b^\dagger_{1,j} b_{0,j} + \sum_{\langle ij \rangle} \frac{1}{z}B_3 b^\dagger_{1,i} b_{1,i} b^\dagger_{2,j} b_{0,j} + \sum_{\langle ij \rangle}\frac{1}{z}C_3 b^\dagger_{1,i} b_{0,i} b^\dagger_{2,j} b_{1,j} + \sum_{\langle ij \rangle} \frac{1}{z}D_3 b^\dagger_{1,i} b_{0,i} b^\dagger_{1,j} b_{2,j}  \nonumber \\
+&\sum_{\langle ij \rangle}\frac{1}{z}E_3 b^\dagger_{0,i} b_{2,i} b^\dagger_{2,j} b_{1,j} +\sum_{\langle ij \rangle}\frac{1}{z}F_3 b^\dagger_{0,i} b_{2,i} b^\dagger_{1,j} b_{2,j} + {\rm H.c.}, \nonumber \\
{\cal H}^{(4)}_{\rm eff}=&\sum_{\langle ij \rangle}\frac{1}{2z}A_4b^\dagger_{1,i} b_{1,i}b^\dagger_{1,j} b_{1,j} + \sum_{\langle ij \rangle}\frac{1}{z}B_4b^\dagger_{1,i} b_{2,i}b^\dagger_{1,j} b_{2,j} + \sum_{\langle ij \rangle}\frac{1}{z}C_4 b^\dagger_{1,i} b_{2,i}b^\dagger_{1,j} b_{1,j} + \sum_{\langle ij \rangle}\frac{1}{2z}D_4b^\dagger_{1,i} b_{2,i}b^\dagger_{2,j} b_{1,j} + {\rm H.c.}, \nonumber
\end{align}
\end{widetext}
where the coefficients such as $A_0,{\tilde A}_{0},A_{1},B_{1},\cdots$ depend on $J$ and $\mu$ via the variational parameter $\theta_{\rm mf}$. The explicit forms are summarized in Appendix \ref{App: coefficients}. 

\subsection{Holstein--Primakoff expansion}\label{Sec: HPE}

In Sec. \ref{Sec: Excitations} we have discussed the fluctuations on the mean-field ground state. In this subsection, we derive the {\it spin-wave} Hamiltonian describing interactions between the Higgs-amplitude and NG-phase modes.

Let us assume that the mean-field approximation is adequate for describing the superfluid state near the Mott-insulator transition. Then we can simplify the Hamiltonian (\ref{eq: effective hamiltonian}) by means of the Holstein--Primakoff expansion \cite{Holstein_1940}. Since the mean-field ground state can be regarded as a Bose--Einstein condensate of the constrained boson $b_{0,i}$, we can eliminate $b_{0,i}$ by an expansion with respect to the fluctuations ({\it spin waves}) $b_{1,i}$ and $b_{2,i}$
\begin{align}
b^\dagger_{m,i}b_{0,j}&=b^\dagger_{m,i}\sqrt{1-b^\dagger_{1,j}b_{1,j}-b^\dagger_{2,j}b_{2,j}}, \label{eq: HP expansion} \\
&\approx b^\dagger_{m,i}-\frac{1}{2}b^\dagger_{m,i}b^\dagger_{1,j}b_{1,j}-\frac{1}{2}b^\dagger_{m,i}b^\dagger_{2,j}b_{2,j}+\cdots.\nonumber
\end{align}
Eliminating $b^\dagger_{0,i}b_{0,i}$ in the Hamiltonian (\ref{eq: effective hamiltonian}) by using the constraint (\ref{eq: constraint2}), and substituting the Holstein--Primakoff expansion (\ref{eq: HP expansion}) into the Hamiltonian (\ref{eq: effective hamiltonian}), we obtain the following series
\begin{align}
{\cal H}_{\rm eff} &\approx {\cal H}^{(0)}_{\rm SW} +  {\cal H}^{(1)}_{\rm SW} +  {\cal H}^{(2)}_{\rm SW} +  {\cal H}^{(3)}_{\rm SW} + {\cal H}^{(4)}_{\rm SW}  \cdots, \label{eq: sw hamiltonian}
\end{align}
where each term ${\cal H}^{(l)}_{\rm SW}$ (for $l=0,1,2,3,4,\cdots$) describes processes involving $l$ collective-mode operators. The control parameter of the Holstein--Primakoff expansion is characterized by the inverse of the spin magnitude $S$. In fact, each term ${\cal H}^{(l)}_{\rm SW}$ is of order $O(S^{2-l/2})$. In this work, in order to evaluate the lowest order effects on the response to the $J$ and $U$ modulations, we deal with fluctuation effects on the response functions up to order $O(S^{0})$. Hence, the expansion (\ref{eq: sw hamiltonian}) is stopped at $l=4$. A similar analysis of another quantum spin system has been made in Ref. \cite{Chernyshev_2009}.

\subsubsection{Subsequent terms of the Holstein--Primakoff expansion}

Let us explain details of the terms ${\cal H}^{(0)}_{\rm SW}$, ${\cal H}^{(1)}_{\rm SW}$, ${\cal H}^{(2)}_{\rm SW}$, and ${\cal H}^{(3)}_{\rm SW}$ in the Holstein--Primakoff expansion (\ref{eq: sw hamiltonian}), respectively. First of all, the zeroth-order term ${\cal H}^{(0)}_{\rm SW}$ is equal to the ground state energy with no fluctuation 
\begin{align}
{\cal H}^{(0)}_{{\rm SW}} 
&= N(A_{0} + {\tilde A}_{0}) \nonumber  \\
&= N E^{\rm MF}(\theta_{\rm mf}), \nonumber 
\end{align}
where $E^{\rm MF}(\theta_{\rm mf})$ is the mean-field energy (\ref{eq: energy}) of the ground state (see Sec.~\ref{Sec: MFGS}).

Next, the linear term ${\cal H}^{(1)}_{\rm SW} = O(S^{3/2})$ is given by
\begin{align}
{\cal H}^{(1)}_{\rm SW} 
& =\sqrt{N}(A_1 + {\tilde A}_1)(b^\dagger_{1,{\bf 0}} + b_{1,{\bf 0}})  \nonumber \\
& \;\;\;\;\;\;\;\;\;\; + \sqrt{N}(B_1 + {\tilde B}_1)(b^\dagger_{2,{\bf 0}} + b_{2,{\bf 0}}),  
\end{align}
where we have introduced the Fourier transformation of the fluctuation operators $b_{1,i}$ and $b_{2,i}$
\begin{align}
b_{m,i}^{\dagger}=\frac{1}{\sqrt{N}}\sum_{{\bf k}\in \Lambda_0} b_{m,{\bf k}}^{\dagger}e^{-{\rm i}{\bf k} \cdot {\bf r}_i},\;\;\;m\in\{1,2\}. \nonumber
\end{align}
The notation $\sum_{{\bf k}\in \Lambda_0}$ denotes that the momentum ${\bf k}$ runs over the cubic-shaped first Brillouin zone $\Lambda_0 \equiv [-\pi,\pi]^3$. For the mean-field ground state, we can easily verify that ${\cal H}^{(1)}_{\rm SW} = 0$.

The quadratic term ${\cal H}^{(2)}_{\rm SW} = O(S)$ can be written as a matrix form
\begin{align}
{\cal H}^{(2)}_{\rm SW}
= \delta E_{2} + \sum_{\lambda=1}^{4}\sum_{\nu=1}^{4}\sum_{{\bf k}\in \Lambda_0} b^\dagger_{\lambda,\bf k} ({\rm H}_{\bf k})_{\lambda\nu} b_{\nu,\bf k}, \label{eq: SWH2}
\end{align}
where $\vec{b}_{\bf k}=(b_{1,{\bf k}},b_{2,{\bf k}},b_{3,{\bf k}},b_{4,{\bf k}})^{\rm T}$ and $(b_{3,{\bf k}},b_{4,{\bf k}}) = (b^{\dagger}_{1,-{\bf k}},b^{\dagger}_{2,-{\bf k}})$. The four dimensional square matrix ${\rm H}_{\bf k}$ is
\begin{align}
{\rm H}_{\bf k} = 
\begin{pmatrix}
f_{11}({\bf k}) & f_{12}({\bf k}) & g_{11}({\bf k}) & g_{12}({\bf k}) \\
f_{21}({\bf k}) & f_{22}({\bf k}) & g_{21}({\bf k}) & g_{22}({\bf k})  \\
g_{11}({\bf k}) & g_{12}({\bf k}) & f_{11}({\bf k}) & f_{12}({\bf k}) \\
g_{21}({\bf k}) & g_{22}({\bf k}) & f_{21}({\bf k}) & f_{22}({\bf k}) \\
\end{pmatrix}. 
\end{align}
The matrix elements of ${\rm H}_{\bf k}$ are given by
\begin{align}
f_{11}({\bf k}) &= (A_2 + {\tilde A}_2 - 2A_0 -{\tilde A}_0 + E_2 \gamma_{\bf k})/2, \nonumber \\
f_{12}({\bf k}) &= f_{21}({\bf k}) = (B_2 + {\tilde B}_2 + F_2 \gamma_{\bf k})/2,  \nonumber \\
f_{22}({\bf k}) &= ({\tilde C}_2 - 2A_{0} - {\tilde A}_0 + I_2 \gamma_{\bf k})/2, \nonumber \\
g_{11}({\bf k}) &= D_2 \gamma_{\bf k}, \nonumber \\
g_{12}({\bf k}) &= g_{21}({\bf k}) = G_2 \gamma_{\bf k}/2, \nonumber \\
g_{22}({\bf k}) &= H_2 \gamma_{\bf k}, \nonumber
\end{align}
where $\gamma_{\bf k}= ({\rm cos}k_x + {\rm cos}k_y + {\rm cos}k_z)/3$ is the band structure of a single particle in the cubic lattice.  At $n_0 \gg 1$, $f_{12}({\bf k})=f_{21}({\bf k})=g_{12}({\bf k})=g_{21}({\bf k})=0$. Thus, ${\cal H}^{(2)}_{\rm SW}$ has no mixing term such as $b^{\dagger}_{1,{\bf k}}b_{2,{\bf k}}$, and we can treat each part labeled by 1 or 2 as an independent branch on each other. This feature stems from the particle-hole symmetry of the effective pseudospin-one model for $n_{0} \gg 1$. In practice, terms with an odd number of $b_{2,{\bf k}}$ are forbidden by the particle-hole symmetry because an exchange between a particle $t_{1,i}$ and hole $t_{-1,i}$ leads to a change of the sign of $b_{2,{\bf k}}$ while such a transformation remains the sign of $b_{1,{\bf k}}$. On the other hand, at lower filling rates, there is no reason that the mixing terms disappear.

Note that we can regard the constant part $\delta E_{2} = -\sum_{{\bf k} \in \Lambda_0}[f_{11}({\bf k}) + f_{22}({\bf k})]$ as a quantum fluctuation correction to the mean-field energy of the ground state ${\cal H}^{(0)}_{\rm SW}(\theta_{\rm mf},\chi_{\rm mf})$. The detailed discussion will be presented in Sec.~\ref{Sec: normal ordering}.

The cubic term ${\cal H}^{(3)}_{\rm SW}$ can be also written as a simple form
\begin{align}
{\cal H}^{(3)}_{\rm SW}
= \frac{1}{\sqrt{N}} \prod_{i=1}^{3} \sum_{\lambda_i = 1}^{4} \sum_{{\bf k}_i \in \Lambda_0} {\rm C}^{(\lambda_1\lambda_2\lambda_3)}_{p_{\lambda_1}{\bf k}_1, p_{\lambda_2}{\bf k}_2, p_{\lambda_3}{\bf k}_3}  \nonumber \\
\times \delta_{{\bf k}_1 + {\bf k}_2 + {\bf k}_3,{\bf 0}} \; b_{\lambda_1, {\bf k}_1} b_{\lambda_2, {\bf k}_2}\ b_{\lambda_3, {\bf k}_3},
\end{align}
where $p_\lambda = 1$ (for $\lambda = 1, 2$) or $p_\lambda = -1$ (for $\lambda = 3, 4$). In addition, $\delta_{{\bf k}_1 + {\bf k}_2 + {\bf k}_3,{\bf 0}}$ is the momentum conservation law satisfied under scattering processes among the three spin waves. The coefficients of the vertices $ {\rm C}^{(\lambda_1\lambda_2\lambda_3)}_{{\bf k}_1, {\bf k}_2, {\bf k}_3} $ which characterize properties of the scattering of the spin wave are given by 
\begin{align}
{\rm C}^{(331)}_{{\bf k}_1,{\bf k}_2,{\bf k}_3} &= {\rm C}^{(131)}_{{\bf k}_1,{\bf k}_2,{\bf k}_3} = (A_{3} - A_{1})\gamma_{{\bf k}_1}, \nonumber \\
{\rm C}^{(342)}_{{\bf k}_1,{\bf k}_2,{\bf k}_3} &= {\rm C}^{(142)}_{{\bf k}_1,{\bf k}_2,{\bf k}_3} = - A_{1}\gamma_{{\bf k}_1},\nonumber \\
{\rm C}^{(431)}_{{\bf k}_1,{\bf k}_2,{\bf k}_3} &= {\rm C}^{(231)}_{{\bf k}_1,{\bf k}_2,{\bf k}_3} = (B_{3} - B_{1})\gamma_{{\bf k}_1}, \nonumber \\
{\rm C}^{(442)}_{{\bf k}_1,{\bf k}_2,{\bf k}_3} &= {\rm C}^{(242)}_{{\bf k}_1,{\bf k}_2,{\bf k}_3} = - B_{1}\gamma_{{\bf k}_1}, \nonumber \\
{\rm C}^{(341)}_{{\bf k}_1,{\bf k}_2,{\bf k}_3} &= {\rm C}^{(132)}_{{\bf k}_1,{\bf k}_2,{\bf k}_3} = C_{3}\gamma_{{\bf k}_1}, \nonumber \\
{\rm C}^{(332)}_{{\bf k}_1,{\bf k}_2,{\bf k}_3} &= {\rm C}^{(141)}_{{\bf k}_1,{\bf k}_2,{\bf k}_3} = D_{3}\gamma_{{\bf k}_1}, \nonumber \\
{\rm C}^{(432)}_{{\bf k}_1,{\bf k}_2,{\bf k}_3} &= {\rm C}^{(241)}_{{\bf k}_1,{\bf k}_2,{\bf k}_3} = E_{3}\gamma_{{\bf k}_1}, \nonumber \\
{\rm C}^{(441)}_{{\bf k}_1,{\bf k}_2,{\bf k}_3} &= {\rm C}^{(232)}_{{\bf k}_1,{\bf k}_2,{\bf k}_3} = F_{3}\gamma_{{\bf k}_1}, \nonumber 
\end{align}
and the others are identically zero. At $n_0 \gg 1$, ${\rm C}^{(431)}_{{\bf k}_1,{\bf k}_2,{\bf k}_3} = {\rm C}^{(231)}_{{\bf k}_1,{\bf k}_2,{\bf k}_3} = {\rm C}^{(442)}_{{\bf k}_1,{\bf k}_2,{\bf k}_3} = {\rm C}^{(242)}_{{\bf k}_1,{\bf k}_2,{\bf k}_3} = {\rm C}^{(341)}_{{\bf k}_1,{\bf k}_2,{\bf k}_3} = {\rm C}^{(132)}_{{\bf k}_1,{\bf k}_2,{\bf k}_3} = {\rm C}^{(332)}_{{\bf k}_1,{\bf k}_2,{\bf k}_3} = {\rm C}^{(141)}_{{\bf k}_1,{\bf k}_2,{\bf k}_3} = 0$ because each vertex characterized by the corresponding coefficient has an odd number of $b_{2,{\bf k}}$.

Finally, we mention a note on the quartic term ${\cal H}^{(4)}_{\rm SW}$. In this work, the quartic term does not enter into the practical analysis of the response functions. The reason is explained in Sec.~\ref{Sec: normal ordering}.

\subsection{Bogoliubov transformation}\label{Sec: Bogoliubov}

In the previous section we have performed the Holstein--Primakoff expansion of the Hamiltonian (\ref{eq: effective hamiltonian}) and discussed properties of each part ${\cal H}^{(l)}_{\rm SW}$. In this section we discuss a Bogoliubov transformation of the quadratic part of the spin-wave Hamiltonian ${\cal H}^{(2)}_{\rm SW}$ and consider the resulting transformation of ${\cal H}^{(3)}_{\rm SW}$. 

Let us define a Bogoliubov transformation 
\begin{align}
\vec{b}_{\bf k} = {\rm W}_{\bf k} \vec{\beta}_{\bf k},\;\;\;\; \vec{\beta}_{\bf k} = {\rm W}^{-1}_{\bf k} \vec{b}_{\lambda,{\bf k}}, \label{eq: bogoliubov}
\end{align}
where $\vec{\beta}_{\bf k}=(\beta_{1,{\bf k}},\beta_{2,{\bf k}},\beta_{3,{\bf k}},\beta_{4,{\bf k}})^{\rm T}$, $\beta_{3,{\bf k}} \equiv \beta^{\dagger}_{1,-{\bf k}}$, $\beta_{4,{\bf k}} \equiv \beta^{\dagger}_{2,-{\bf k}}$, $[\beta_{m,{\bf k}},\beta^{\dagger}_{n,{\bf k}'}] = \delta_{m,n}\delta_{{\bf k},{\bf k}'}$ (for $m,n = 1,2$), and $[\beta_{m,{\bf k}},\beta_{n,{\bf k}'}] = [\beta^\dagger_{m,{\bf k}},\beta^\dagger_{n,{\bf k}'}] = 0$. In general, the matrix elements of ${\rm W}_{\bf k}$ can be written as
\begin{align}
{\rm W}_{\bf k} = 
\begin{pmatrix}
u_{11}({\bf k}) & u_{12}({\bf k}) & v_{11}({\bf k})  & v_{12}({\bf k})  \\
u_{21}({\bf k}) & u_{22}({\bf k}) & v_{21}({\bf k}) & v_{22}({\bf k}) \\
v^*_{11}(-{\bf k})   & v^*_{12}(-{\bf k})    & u^*_{11}(-{\bf k}) & u^*_{12}(-{\bf k})   \\
v^*_{21}(-{\bf k})   & v^*_{22}(-{\bf k})   & u^*_{21}(-{\bf k}) & u^*_{22}(-{\bf k})  
\end{pmatrix}. 
\end{align}
The transformation ${\rm W}_{\bf k}$ fulfills a condition
\begin{align}
{\rm W}_{\bf k} {\rm g} {\rm W}^{\dagger}_{\bf k} = {\rm W}^{\dagger}_{\bf k} {\rm g} {\rm W}_{\bf k} = {\rm g}, \label{eq: bogoliubov condition1}
\end{align}
where ${\rm g} ={\rm diag}(1,1,-1,-1)$ is the metric tensor in the Minkowski space ${\mathbb M}^{2 \otimes 2}$, because of the Bose statistics of the new operators. In addition, in order to diagonalize ${\cal H}^{(2)}_{\rm SW}$ so that
\begin{align}
{\cal H}^{(2)}_{\rm SW} 
=  \delta E_{2} + \sum_{\lambda=1}^{4} \sum_{\nu=1}^{4} \sum_{{\bf k} \in \Lambda_0} \beta^{\dagger}_{\lambda,{\bf k}} ({\rm D}_{\bf k})_{\lambda \nu} \beta_{\nu,{\bf k}}, \label{eq: bogoliubov diagonal}
\end{align}
where ${\rm D}_{\bf k} = {\rm diag}(e_1({\bf k}),e_2({\bf k}),e_3({\bf k}),e_4({\bf k}))$ is a diagonal matrix, we impose on the matrix ${\rm W}_{\bf k}$ a condition 
\begin{align}
{\rm W}_{\bf k}^{-1} ({\rm g}{\rm H}_{\bf k}) {\rm W}_{\bf k} = {\rm g}{\rm D}_{\bf k}. \label{eq: bogoliubov condition2}
\end{align}
Solving the eigenvalue problem of the non-Hermite matrix ${\rm g}{\rm H}_{\bf k}$ defined by Eqs (\ref{eq: bogoliubov condition1}) and (\ref{eq: bogoliubov condition2}), we obtain the specific form of ${\rm W}_{\bf k}$ and dispersion relations ${\cal E}_{1,{\bf k}}=e_1({\bf k}) + e_3({\bf k})$ of $\beta_{1,{\bf k}}$ and ${\cal E}_{2,{\bf k}}=e_2({\bf k}) + e_4({\bf k})$ of $\beta_{2,{\bf k}}$. Notice that the dispersions ${\cal E}_{1,{\bf k}}$ and ${\cal E}_{2,{\bf k}}$ correspond to the Higgs and NG modes, respectively. For more details of the eigenvalue problem of ${\rm g}{\rm H}_{\bf k}$, see Ref. \cite{Huber_2007}.

At $n_0 \gg1$, the different sectors labeled by 1 or 2 are completely decoupled, so that we can easily diagonalize ${\rm gH_{\rm k}}$ and obtain the dispersion relations of the collective modes and coefficient matrix ${\rm W}_{\bf k}$ as analytical forms. In Appendix \ref{App: bogoliubov}, we will demonstrate it in practice. On the other hand, at lower filling rates, to compute ${\rm W}_{\bf k}$ and the dispersion relations is possible but more complicated than the large filling case. In this paper, we calculate them by a numerical diagonalization of the non-Hermite matrix ${\rm g}{\rm H}_{\bf k}$. The analytic expressions of the dispersion relations in the superfluid phase at an arbitrary filling rate have been obtained in Ref. \cite{Huber_2007}.

After the Bogoliubov transformation of the quadratic part ${\cal H}^{(2)}_{\rm SW}$, the cubic term ${\cal H}^{(3)}_{\rm SW}$ becomes 
\begin{align}
{\cal H}^{(3)}_{\rm SW} 
= \frac{1}{\sqrt{N}}\prod_{i=1}^{3} \sum^{4}_{\lambda_i = 1} \sum_{{\bf k}_i \in \Lambda_0} {\rm M}^{(\lambda_1\lambda_2\lambda_3)}_{p_{\lambda_1}{\bf k}_1,p_{\lambda_2}{\bf k}_2,p_{\lambda_3}{\bf k}_3} \nonumber  \\
\times \delta_{{\bf k}_1 + {\bf k}_2 + {\bf k}_3,{\bf 0}} \; \beta_{\lambda_1,{\bf k}_1}\beta_{\lambda_2,{\bf k}_2}\beta_{\lambda_3,{\bf k}_3}. 
\end{align}
Here, the new coefficients ${\rm M}^{(\lambda_1\lambda_2\lambda_3)}_{{\bf k}_1,{\bf k}_2,{\bf k}_3}$ are related to ${\rm C}^{(\lambda_1\lambda_2\lambda_3)}_{{\bf k}_1, {\bf k}_2, {\bf k}_3}$ by a relation 
\begin{align}
{\rm M}^{(\lambda_1\lambda_2\lambda_3)}_{p_{\lambda_1}{\bf k}_1, p_{\lambda_2}{\bf k}_2, p_{\lambda_3}{\bf k}_3}
&=\sum_{\nu_1,\nu_2,\nu_3 = 1}^{4} {\rm C}^{(\nu_1\nu_2\nu_3)}_{p_{\nu_1}{\bf k}_1,p_{\nu_2}{\bf k}_2,p_{\nu_3}{\bf k}_3} \nonumber \\
&\times ({\rm W}_{{\bf k}_1})_{\nu_1}^{\;\; \lambda_1}({\rm W}_{{\bf k}_2})_{\nu_2}^{\;\; \lambda_2}({\rm W}_{{\bf k}_3})_{\nu_3}^{\;\; \lambda_3}. 
\end{align}

The coefficients ${\rm M}^{(\lambda_1\lambda_2\lambda_3)}_{{\bf k}_1,{\bf k}_2,{\bf k}_3}$ characterize the interactions among the three collective modes of the diagonalized basis. For $n_0 \gg 1$, processes with an odd number of the NG modes are prohibited due to the particle-hole symmetry of the effective pseudospin-one model (\ref{eq: high filling efm}). On the other hand, the effective model at lower filling rates has no longer such a symmetry, thus, permits not only the even-NG processes but also the odd-NG processes. As we will see in Sec.~\ref{Sec: normal} in contrast to the large filling case, new types of contribution to the response properties emerge due to the physical background.

\subsection{Normal ordering}\label{Sec: normal ordering}

So far we have discussed the Bogoliubov transformation of the spin-wave Hamiltonian ${\cal H}_{\rm SW}$. Obviously, the resulting Hamiltonian after the Bogoliubov transformation is not normally ordered with respect to the Bogoliubov operators $\beta_{m,{\bf k}}$. In Sec. \ref{Sec: LRA}, we will apply the field theoretical methods for the spin-wave Hamiltonian in order to calculate the response functions. Therefore, it is necessary to obtain a normally ordered form of ${\cal H}_{\rm SW}$. 

Let us consider the normal ordering of the quadratic part of the spin-wave Hamiltonian, ${\cal H}_{\rm SW}^{(2)}$. In the following discussion, a notation $: \cdot :$ represents a normal ordered operator with respect to the Bogoliubov operators. In the quadratic Hamiltonian, each out of normally ordered terms produces a constant shift after permutations between the canonical operators $\beta_{m,{\bf k}}$ and $\beta^\dagger_{m,{\bf k}}$. Thus the quadratic Hamiltonian can be rewritten as
\begin{align}
{\cal H}_{\rm SW}^{(2)} 
&= \delta E_{2} + \delta {\tilde E}_{2}+ :{\tilde {\cal H}}_{\rm SW}^{(2)}:\;, \nonumber
\end{align}
where $\delta {\tilde E}_{2}$ is the resulting constant shift arising after ${\tilde {\cal H}}_{\rm SW}^{(2)}$ is normally ordered. In a similar way, the cubic and quartic Hamiltonians, ${\cal H}_{\rm SW}^{(3)}$ and ${\cal H}_{\rm SW}^{(4)}$, become
\begin{align}
{\cal H}_{\rm SW}^{(3)} 
&= :{\cal H}_{\rm SW}^{(3)}: + \delta {\cal H}_{\rm SW}^{(1)}, \nonumber \\
{\cal H}_{\rm SW}^{(4)} 
&= \delta E_{4} + :\delta {\cal H}_{\rm SW}^{(2)}: + :{\cal H}_{\rm SW}^{(4)}:\;, \nonumber 
\end{align}
where $\delta {\cal H}_{\rm SW}^{(1)}$, $\delta E_{4}$, and $:\delta {\cal H}_{\rm SW}^{(2)}:$ are the resulting linear, constant, and quadratic shifts arising after making the cubic and quartic Hamiltonians normally ordered.

The total shift $\delta E_2 + \delta {\tilde E_2} + \delta E_4$ can be interpreted as a fluctuation correction to the mean-field energy of the ground state ${\cal H}^{(0)}_{\rm SW}$ \cite{Huber_2007}. The first two terms represent $1/S$ corrections to the ground-state energy and the last term is a higher order correction of order $1/S^2$. To minimize the modified ground-state energy with respect to $\theta$ and $\chi$ leads to a renormalization of the variational parameters of the mean-field configuration: $\theta_{\rm mf} \rightarrow \theta_{\rm ren} = \theta_{\rm mf} + \delta \theta_{\rm cor}$ and $\chi_{\rm mf} \rightarrow \chi_{\rm ren} = \chi_{\rm mf} + \delta \chi_{\rm cor}$. This corresponds to a reduction of the order-parameter amplitude induced by quantum and thermal fluctuations. At the renormalized configuration, the linear term including the shift from the cubic Hamiltonian ${\cal H}^{(3)}_{\rm SW}$ becomes zero: ${\cal H}^{(1)}_{\rm SW} + \delta {\cal H}^{(1)}_{\rm SW} = 0$. Moreover, the renormalized parameters and additional quadratic term $:\delta {\cal H}_{\rm SW}^{(2)}:$ stemming from the quartic Hamiltonian ${\cal H}^{(4)}_{\rm SW}$ modify the band energies ${\cal E}_{1,{\bf k}}$ and ${\cal E}_{2,{\bf k}}$ calculated within the mean-field approximation.
 
Although it is naively expected that inclusion of the renormalization effect induced by fluctuations should make the results more quantitative, it leads to a theoretical difficulty concerned with spectral properties of the NG mode. If we deal with the renormalization effect on the basis of our perturbative scheme around the mean-field ground state, then we are confronted with a situation in which a finite energy gap opens in the NG mode branch. In general, the gap of the NG mode must vanish in the symmetry broken phase, so that the appearance of the finite gap is an artifact of our naive perturbative approach. Moreover, whether the finite gap exists or not in the low energy sector of the NG mode spectrum strongly affects the decay processes of the Higgs mode because the possible scattering channels are restricted by the on-shell energy-momentum conservation laws between the low-energy collective modes \cite{Nagao_2016}. Thus, in order to describe the stability of the Higgs mode corresponding to experiments, we need to eliminate the finite gap from the NG mode branch.

The similar problem also appears in the Hartree-Fock-Bogoliubov approximation of single component dilute Bose gases \cite{Griffin_1996, Kita_2006, Yukalov_2006, Pethick_2008}. In this scheme, the artificial energy gap of the NG or Bogoliubov mode is often eliminated by the conventional Popov--Shohno prescription \cite{Pethick_2008, Shohno_1964, Griffin_1996} in which an anomalous average of boson operators is detuned so that the artificial gap vanishes. Our bosons in the current problem have two components, so that the application of the similar prescription for our case is not straightforward. Therefore, in this work, we do neglect the modification of the mean-field variational parameters as a simpler prescription. Our prescription here is similar in spirit to the standard Bogoliubov approximation for dilute Bose gases \cite{Pitaevskii_2003}, and is expected to be better as the spatial dimension of the system increases and the temperature decreases.

In addition to the prescription, we also neglect the normally ordered quartic term $:{\cal H}_{\rm SW}^{(4)}:$ throughout our analysis. Within our lowest order $O(S^0)$, the term only generates a shift of the peak position of the Higgs mode but no contribution to the peak width. Moreover, the shift is expected to be rather small at sufficiently low temperatures. Thus it makes no important difference whether the quartic term exists or not, as far as the problem of the stability of the Higgs mode is concerned.

Finally, the above discussions are summarized as the following normally ordered Hamiltonian: 
\begin{align}
{\cal H}_{\rm SW} = {\rm const}. + :{\tilde {\cal H}}^{(2)}_{\rm SW}: + :{\cal H}^{(3)}_{\rm SW}:. \label{eq: final hamiltonian}
\end{align}
In the next section, we will compute the fluctuation corrections to the response functions practically by using the final Hamiltonian (\ref{eq: final hamiltonian}). 

\section{Linear response analysis}\label{Sec: LRA}

In this section, we calculate and investigate the response functions (\ref{eq: response_kk}) and (\ref{eq: response_oo}) combining the methods developed in Sec. \ref{Sec: Methods} and imaginary-time or Matsubara Green's function theory. The basis of the Green's function theory is explained in Ref. \cite{Abrikosov_1975, Lifshitz_1980,Altland_2010}.

\subsection{Response functions}\label{Sec: response functions}
We express the $K$-to-$K$ response function (\ref{eq: response_kk}) in terms of the Bogoliubov operators $\beta_{m,{\bf k}}$. Using the Holstein--Primakoff expansion (\ref{eq: HP expansion}) and Bogoliubov transformation, the kinetic energy $K$ becomes
\begin{align}
K = NA_0 + \sqrt{N} \Upsilon_1 (\beta^\dagger_{1,{\bf 0}} + \beta_{1,{\bf 0}}) +  \cdots, \label{eq: hpe_k}
\end{align}
where the coefficient $\Upsilon_1$ is defined by 
\begin{align}
\Upsilon_1 = A_1[u_{11}({\bf 0}) + v_{11}({\bf 0})] + B_1[u_{21}({\bf 0}) + v_{21}({\bf 0})]. \label{eq: coefficient 1}
\end{align}
It should be noted that $\cdots$ in Eq.~(\ref{eq: hpe_k}) includes the term proportional to $\beta_{2,{\bf 0}} + \beta^\dagger_{2,{\bf 0}}$. Here $\beta_{2,{\bf 0}}$ corresponds to the zero energy mode of the system. We can easily check that the coefficient of $\beta_{2,{\bf 0}} + \beta^\dagger_{2,{\bf 0}}$ should be zero because of the eigenvalue equation of $[u_{12}({\bf 0}), v_{12}({\bf 0}), u_{22}({\bf 0}), v_{22}({\bf 0})]$ extracted from Eq.~(\ref{eq: bogoliubov condition2}).

Therefore the zero mode contribution is eliminated in our analysis. Substituting (\ref{eq: hpe_k}) into the definition of $D^{\rm R}_{KK}(t-t')$ and keeping only leading order terms, we obtain
\begin{align}
D_{KK}^{\rm R}(t-t') 
&= N |\Upsilon_1|^2 \left\{ G^{\rm R}_{13,{\bf 0}}(t-t')  +G^{\rm R}_{31,{\bf 0}}(t-t')  \right. \nonumber \\
& \left. + G^{\rm R}_{11,{\bf 0}}(t-t')  + G^{\rm R}_{33,{\bf 0}}(t-t')  \right\}, \label{eq: response_kk bogoliubov}
\end{align}
where we have introduced four types of retarded Green's function of the zero-momentum Higgs mode $\beta_{1,{\bf 0}}$,
\begin{align}
G^{\rm R}_{13,{\bf 0}}(t-t')&=-i\Theta(t-t') \langle [\beta_{1,{\bf 0}}(t), \beta^\dagger_{1,{\bf 0}}(t')] \rangle_{\rm eq}, \nonumber \\
G^{\rm R}_{11,{\bf 0}}(t-t')&=-i\Theta(t-t') \langle [\beta_{1,{\bf 0}}(t),\beta_{1,{\bf 0}}(t')] \rangle_{\rm eq},  \nonumber \\
G^{\rm R}_{31,{\bf 0}}(t-t')&=-i\Theta(t-t') \langle [\beta^\dagger_{1,{\bf 0}}(t), \beta_{1,{\bf 0}}(t')] \rangle_{\rm eq},  \nonumber \\
G^{\rm R}_{33,{\bf 0}}(t-t')&=-i\Theta(t-t') \langle [\beta^\dagger_{1,{\bf 0}}(t),\beta^\dagger_{1,{\bf 0}}(t')] \rangle_{\rm eq}. \nonumber
\end{align}
Thus, up to the leading order, evaluating the response function $D_{KK}^{\rm R}(t-t')$ results in calculating these retarded functions of the Higgs mode. The Fourier transformation with respect to $t-t'$ gives the dynamical susceptibility $\chi_{KK}(\omega)$, which characterizes the stability of the Higgs mode [see Eq. (\ref{eq: susceptibility_kk})].

Similarly, we can obtain the $O$-to-$O$ response function (\ref{eq: response_oo}) written in terms of the Bogoliubov operators. Using the Holstein--Primakoff expansion (\ref{eq: HP expansion}) and Bogoliubov transformation, the onsite interaction energy $O$ becomes
\begin{align}
O = N {\tilde A}_0' + \sqrt{N} \Upsilon_2 (\beta^\dagger_{1,{\bf 0}} + \beta_{1,{\bf 0}}) + \cdots, \label{eq: hpe_o}
\end{align}
where the coefficient $\Upsilon_2 $ is defined by
\begin{align}
\Upsilon_2 = {\tilde A}_1' [u_{11}({\bf 0}) + v_{11}({\bf 0})],
\end{align}
and the constants ${\tilde A}_0'$ and ${\tilde A}_1'$ are given by 
\begin{align}
{\tilde A}'_0 = \frac{1}{2}s_{1}^2,\;\;{\tilde A}'_1 = -\frac{1}{2}s_{1}c_{1}.
\end{align}
For the same reason of the zero-mode coefficient vanishing in $K$, the onsite energy $O$ has also no term of $\beta_{2,{\bf 0}}+\beta^{\dagger}_{2,{\bf 0}}$.

Substituting (\ref{eq: hpe_o}) into the definition of $D^{\rm R}_{OO}(t-t')$ and keeping only leading order terms, we obtain
\begin{align}
D_{OO}^{\rm R}(t-t') 
= & N |\Upsilon_2|^2 \left\{ G^{\rm R}_{13,{\bf 0}}(t-t') + G^{\rm R}_{31,{\bf 0}}(t-t') \right. \nonumber \\
&\left. + G^{\rm R}_{11,{\bf 0}}(t-t') + G^{\rm R}_{33,{\bf 0}}(t-t')\right\}.
\end{align}
The dynamical susceptibility $\chi_{OO}(\omega)$ is given by the Fourier transformation of this quantity [see Eq. (\ref{eq: susceptibility_oo})].

\begin{figure}
\includegraphics[width=80mm]{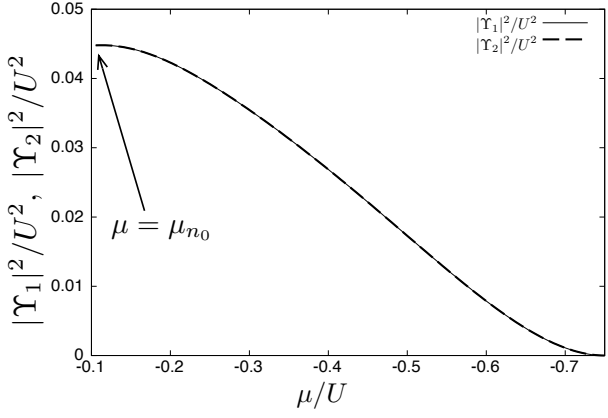}
\vspace{-2mm}
\caption{Chemical potential dependence of the coefficients $|\Upsilon_1|^2$ and $|\Upsilon_2|^2$. We choose the specific parameters $n_0=1$ and $u=1\;(zJ/U=0.25)$. The solid and dashed lines represent $|\Upsilon_1|^2/U^2$ and $|\Upsilon_2|^2/U^2$, which characterize the response magnitude of the hopping and onsite-interaction modulations, respectively. The point indicated by the solid arrow corresponds to the unit filling rate ${\bar n} = n_0=1$. The mean filling rate ${\bar n}$ decreases as the chemical potential $\mu$ decreases. After decreasing below $\mu \approx -0.75$, ${\bar n}$ becomes zero.}
\label{fig: upsilon}
\end{figure}

Within the leading order, the $O$-to-$O$ response function has the same form as the $K$-to-$K$ response function except for its coefficients $|\Upsilon_1|^2$ and $|\Upsilon_2|^2$. In Fig.~\ref{fig: upsilon}, we show the chemical potential dependence of the coefficients at ${\bar n}=n_0=1$ and $u=1$. The point indicated by a solid arrow in Fig.~\ref{fig: upsilon} is at the commensurate filling rate $n_0$, and the corresponding chemical potential is expressed by $\mu_{n_0}$ whose explicit form is presented in Appendix \ref{App: variational}. As shown in Fig.~\ref{fig: upsilon}, the coefficients are found to completely coincide with each other for any $\mu$, so that there is no difference between two response functions, at least, within our approximate calculation. Notice that the similar coincidence occurs for other values of $n_0$ and $u$.

\subsection{Imaginary-time Green's functions}

In this paper, we calculate the response functions by means of perturbative methods of the imaginary-time Green's functions. Let us define three time-ordered {\it normal} or {\it anomalous} Green's functions on an imaginary time axis  \cite{Abrikosov_1975,Lifshitz_1980,Altland_2010}
\begin{align}
G_{1,{\bf k}}(\tau-\tau') &=- \langle T_{\tau} \beta_{1,\bf k}(\tau) \beta^{\dagger}_{1,\bf k}(\tau') \rangle_{\rm eq} \nonumber \\
&\;\;\;\;\;\;\;\;\;\;\;\;\;\;\;\;\;\;\; + \langle \beta_{1,{\bf 0}}(0) \rangle_{\rm eq}\langle \beta^{\dagger}_{1,{\bf 0}}(0) \rangle_{\rm eq}, \nonumber \\
F_{1,{\bf k}}(\tau-\tau') &=-\langle T_{\tau} \beta_{1,{\bf k}}(\tau) \beta_{1,-{\bf k}}(\tau') \rangle_{\rm eq} + \langle \beta_{1,{\bf 0}}(0) \rangle_{\rm eq}^2, \nonumber \\
F^{\dagger}_{1,{\bf k}}(\tau-\tau') &=-\langle T_{\tau} \beta^\dagger_{1,-{\bf k}}(\tau) \beta^\dagger_{1,{\bf k}}(\tau') \rangle_{\rm eq} + \langle \beta^{\dagger}_{1,{\bf 0}}(0) \rangle_{\rm eq}^2. \nonumber
\end{align}
Here $T_{\tau}$ is the imaginary-time ordering operator and $\tau,\tau' \in [0,\beta]$. These functions are periodic with respect to the inverse temperature $\beta$ \cite{Abrikosov_1975,Lifshitz_1980,Altland_2010}. Thus the Fourier components are given by
\begin{align}
{\cal G}_{1,{\bf k}}(i\omega_n) &= \int^{\beta}_{0} d\tau G_{1,{\bf k}}(\tau)e^{i\omega_n\tau}, \nonumber \\
{\cal F}_{1,{\bf k}}(i\omega_n) &= \int^{\beta}_{0} d\tau F_{1,{\bf k}}(\tau)e^{i\omega_n\tau}, \nonumber \\
{\cal F}^{\dagger}_{1,{\bf k}}(i\omega_n) &= \int^{\beta}_{0} d\tau F^{\dagger}_{1,{\bf k}}(\tau)e^{i\omega_n\tau}. \nonumber 
\end{align}
where $\omega_{n}=2\pi n / \beta$ ($n \in {\mathbb N}$) is the Matsubara frequency \cite{Abrikosov_1975,Lifshitz_1980,Altland_2010}. It should be noted that a relation ${\cal F}^{\dagger}_{1,{\bf k}}(i\omega_n)={\cal F}_{1,{\bf k}}(-i\omega_n)$ holds for any $n$, at least, within our leading order $O(S^0)$. In fact, this is verified by a straightforward calculation based on the perturbative expansion. According to more general consideration \cite{Lifshitz_1980}, this relation is expected to be true at any order of the perturbative expansion.

For a fixed $\omega_n$, the imaginary-time Green's functions ${\cal G}(i\omega_n)$ and ${\cal F}(i\omega_n)$ fulfill the Dyson's equations \cite{Abrikosov_1975, Lifshitz_1980}
\begin{align}
{\cal G}_{1,{\bf 0}}(i\omega_n) 
= &\; {\cal G}^{(0)}_{1,{\bf 0}}(i\omega_n) + {\cal G}^{(0)}_{1,{\bf 0}}(i\omega_n) \Sigma_{11}(i\omega_n) {\cal G}_{1,{\bf 0}}(i\omega_n) \nonumber \\
&\;\;\;\;\;\;\;\;\;\;\;\; +  {\cal G}^{(0)}_{1,{\bf 0}}(i\omega_n) \Sigma_{02}(i\omega_n) {\cal F}_{1,{\bf 0}}(-i\omega_n), \nonumber\\
{\cal F}_{1,{\bf 0}}(i\omega_n) 
= &\; {\cal G}^{(0)}_{1,{\bf 0}}(i\omega_n) \Sigma_{11}(i\omega_n) {\cal F}_{1,{\bf 0}}(i\omega_n) \nonumber \\
&\;\;\;\;\;\;\;\;\;\;\;\; + {\cal G}^{(0)}_{1,{\bf 0}}(i\omega_n) \Sigma_{02}(i\omega_n) {\cal G}_{1,{\bf 0}}(-i\omega_n),\nonumber
\end{align}
where $\Sigma_{11}(i\omega_n)$ and $\Sigma_{02}(i\omega_n)$ are the self-energy functions of the normal and anomalous Green's functions. Here, $ {\cal G}^{(0)}_{1,{\bf 0}}(i\omega_n)=1/(i\omega_n - \Delta)$ is the free propagator of the Higgs mode with its energy gap $\Delta$ at zero momentum. The formal solutions \cite{Abrikosov_1975, Lifshitz_1980} are
\begin{align}
{\cal G}_{1,{\bf 0}}(i\omega_n) &= -\frac{1}{D}\left\{ \left[ {\cal G}^{(0)}_{1,{\bf 0}}(-i\omega_n) \right]^{-1} - \Sigma_{11}(-i\omega_n) \right\}, \nonumber \\
{\cal F}_{1,{\bf 0}}(i\omega_n) &= -\frac{1}{D} \Sigma_{02}(i\omega_n), \nonumber 
\end{align}
where its denominator $D$ is given by
\begin{align}
D = & \left[ \Sigma_{02}(i\omega_n) \right]^2 - \left[ i\omega_n - \Delta - \Sigma_{11}(i\omega_n) \right] \nonumber \\
& \;\;\;\;\;\;\;\;\;\;\;\; \times \left[ -i\omega_n - \Delta - \Sigma_{11}(-i\omega_n) \right]. \nonumber  
\end{align}

In terms with the Fourier components of the Green's functions, the dynamical susceptibilities $\chi_{KK}(\omega)$ and $\chi_{OO}(\omega)$ read
\begin{align}
\chi_{KK}(\omega) &= N|\Upsilon_1|^2 \left\{ g(\omega) + f(\omega) + {\bar g}(\omega) + {\bar f}(\omega) \right\}, \label{eq: tilde susceptibility_kk}  \\
\chi_{OO}(\omega) &= N|\Upsilon_2|^2 \left\{ g(\omega) + f(\omega) + {\bar g}(\omega) + {\bar f}(\omega) \right\}. \label{eq: tilde susceptibility_oo}
\end{align}
The analytically continued functions 
\begin{align}
g(\omega) &= {\cal G}_{1,{\bf 0}}(i\omega_n)|_{i\omega_n \rightarrow \omega + i\epsilon},\;\;f(\omega) = {\cal F}_{1,{\bf 0}}(i\omega_n)|_{i\omega_n \rightarrow \omega + i\epsilon}, \nonumber \\
{\bar g}(\omega) &={\cal G}_{1,{\bf 0}}(-i\omega_n)|_{i\omega_n \rightarrow \omega + i\epsilon}, \nonumber \\
{\bar f}(\omega) &={\cal F}^\dagger_{1,{\bf 0}}(i\omega_n)|_{i\omega_n \rightarrow \omega + i\epsilon} = {\cal F}_{1,{\bf 0}}(-i\omega_n)|_{i\omega_n \rightarrow \omega + i\epsilon}, \nonumber
\end{align}
are nothing but the retarded (real time) Green's functions $G^{\rm R}_{13,{\bf 0}}(\omega)$, $G^{\rm R}_{11,{\bf 0}}(\omega)$, $G^{\rm R}_{31,{\bf 0}}(\omega)$, and $G^{\rm R}_{33,{\bf 0}}(\omega)$. Here $\epsilon$ is an infinitesimal and positive number. Thus we can obtain the dynamical susceptibilities when the self-energy functions $\Sigma_{11}(i\omega_n)$ and $\Sigma_{02}(i\omega_n)$ are known for all of $n>0$. 

\subsection{Self-energy functions}\label{Sec: normal}

At the level of the formal solutions of the Dyson's equations, the self-energy functions are still unknown. Here we compute them through a perturbative approximation of the normal and anomalous Green's functions. The lowest order contributions to the self-energy functions arise from the second order perturbation with respect to $:{\cal H}^{(3)}_{\rm SW}:$. If we stop the expansion up to the lowest one-loop order, i.e., $O(S^0)$, each self-energy function then contains twelve number of distinct contributions.

\begin{figure*}
\includegraphics[width=150mm]{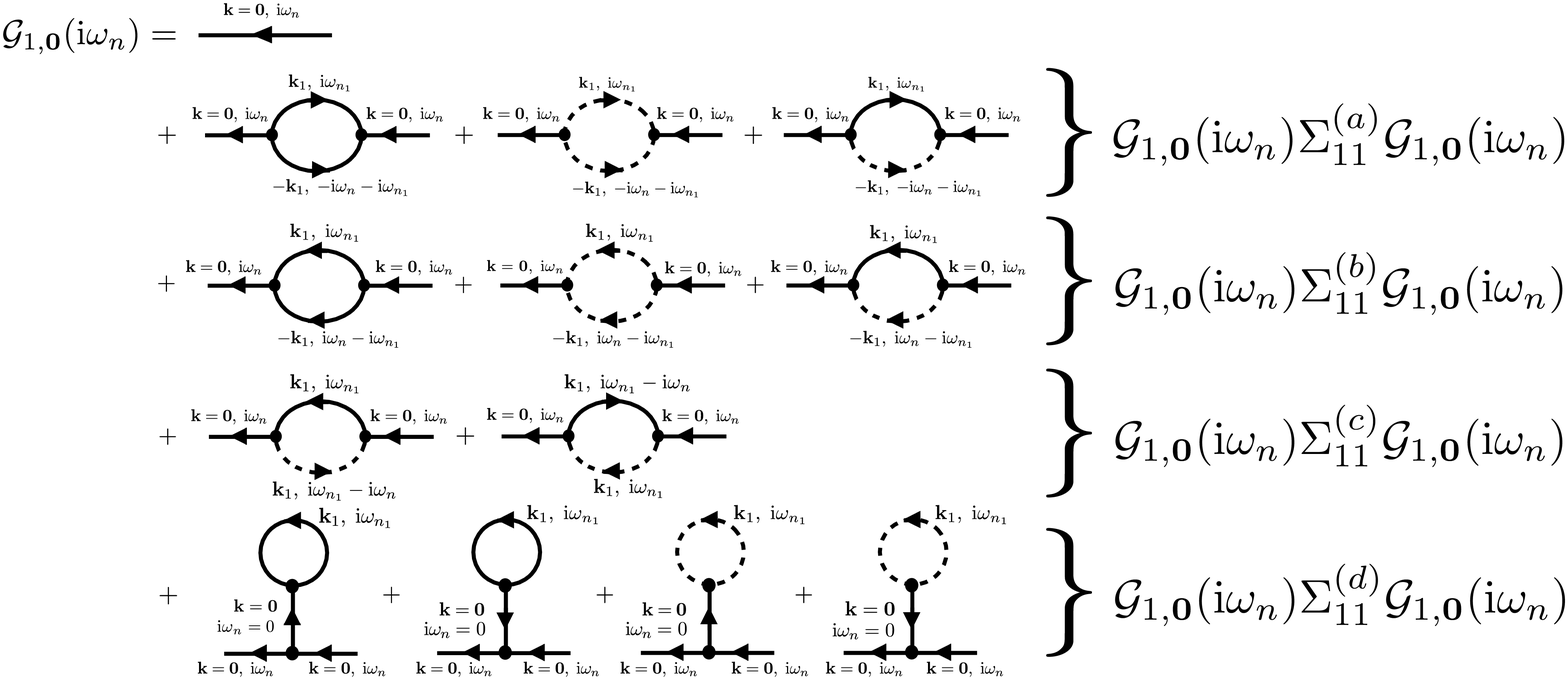}
\vspace{0mm}
\caption{Leading one-loop order contributions to the normal Green's function ${\cal G}_{1,{\bf 0}}(i\omega_n)$. The variables ${\bf k}_1$ and $i\omega_{n_1}$ added near the internal line implies the internal summation over the possible momentum and Matsubara frequency. The solid arrow denotes the propagator of the free Higgs mode while the dashed arrow denotes the propagator of the free NG mode. The self-energy function of each diagram is obtained by picking off its two external lines. The left arrow in the first column represents the zeroth order Green's function. The diagrams in each column form an individual group labeled by $\Sigma^{(a)}_{11}(i\omega_n)$, $\Sigma^{(b)}_{11}(i\omega_n)$, $\Sigma^{(c)}_{11}(i\omega_n)$, or $\Sigma^{(d)}_{11}(i\omega_n)$.}
\label{fig: normal self-energy function}
\end{figure*}

Let us compute the normal self-energy function $\Sigma_{11}(i\omega_n)$. In Fig. \ref{fig: normal self-energy function}, we show the corresponding Feynman diagrams up to the one-loop order corrections. The contributions to the full self-energy function can be categorized into four parts, $\Sigma_{11}(i\omega_n)=\Sigma^{(a)}_{11}(i\omega_n)+\Sigma^{(b)}_{11}(i\omega_n)+\Sigma^{(c)}_{11}(i\omega_n)+\Sigma^{(d)}_{11}(i\omega_n)$. Within the lowest order each partial self-energy function is given by 
\begin{widetext}
\begin{align}
\Sigma^{(a)}_{11}(i\omega_n) &=\frac{-1}{2N}\sum_{{\bf k}_1 \in \Lambda_0} M^{[333]}_{{\bf 0},{\bf k}_1,-{\bf k}_1} M^{[111]}_{{\bf k}_1,-{\bf k}_1,{\bf 0}} \frac{1 + 2n_{\rm B}[{\cal E}_{1,{\bf k}_1}]}{i\omega_n + 2{\cal E}_{1,{\bf k}_1}} -\frac{1}{2N}\sum_{{\bf k}_1 \in \Lambda_0} M^{[344]}_{{\bf 0},{\bf k}_1,-{\bf k}_1} M^{[221]}_{{\bf k}_1,-{\bf k}_1,{\bf 0}} \frac{1 + 2n_{\rm B}[{\cal E}_{2,{\bf k}_1}]}{i\omega_n + 2{\cal E}_{2,{\bf k}_1}}  \nonumber \\
&-\frac{1}{N}\sum_{{\bf k}_1 \in \Lambda_0} M^{[334]}_{{\bf 0},{\bf k}_1,-{\bf k}_1} M^{[211]}_{-{\bf k}_1,{\bf k}_1,{\bf 0}} \frac{1 + n_{\rm B}[{\cal E}_{1,{\bf k}_1}] + n_{\rm B}[{\cal E}_{2,{\bf k}_1}]}{i\omega_n + {\cal E}_{1,{\bf k}_1} + {\cal E}_{2,{\bf k}_1}}, \nonumber \\
\Sigma^{(b)}_{11}(i\omega_n) &= \frac{1}{2N}\sum_{{\bf k}_1 \in \Lambda_0} M^{[311]}_{{\bf 0},{\bf k}_1,-{\bf k}_1} M^{[331]}_{{\bf k}_1,-{\bf k}_1,{\bf 0}} \frac{1 + 2n_{\rm B}[{\cal E}_{1,{\bf k}_1}]}{i\omega_n - 2{\cal E}_{1,{\bf k}_1}} + \frac{1}{2N}\sum_{{\bf k}_1 \in \Lambda_0} M^{[322]}_{{\bf 0},{\bf k}_1,-{\bf k}_1} M^{[441]}_{{\bf k}_1,-{\bf k}_1,{\bf 0}}  \frac{1 + 2n_{\rm B}[{\cal E}_{2,{\bf k}_1}]}{i\omega_n - 2{\cal E}_{2,{\bf k}_1}}    \nonumber \\
&+\frac{1}{N}\sum_{{\bf k}_1 \in \Lambda_0} M^{[312]}_{{\bf 0},{\bf k}_1,-{\bf k}_1} M^{[341]}_{{\bf k}_1,-{\bf k}_1,{\bf 0}}  \frac{1 + n_{\rm B}[{\cal E}_{1,{\bf k}_1}] + n_{\rm B}[{\cal E}_{2,{\bf k}_1}]}{i\omega_n - {\cal E}_{1,{\bf k}_1} - {\cal E}_{2,{\bf k}_1}}, \nonumber \\
\Sigma^{(c)}_{11}(i\omega_n) &= -\frac{1}{N} \sum_{{\bf k}_1 \in \Lambda_0} M^{[332]}_{{\bf 0},{\bf k}_1,{\bf k}_1} M^{[411]}_{{\bf k}_1,{\bf k}_1,{\bf 0}} \frac{n_{\rm B}[{\cal E}_{2,{\bf k}_1}] - n_{\rm B}[{\cal E}_{1,{\bf k}_1}]}{i\omega_n + {\cal E}_{1,{\bf k}_1} - {\cal E}_{2,{\bf k}_1}} -\frac{1}{N} \sum_{{\bf k}_1 \in \Lambda_0} M^{[341]}_{{\bf 0},{\bf k}_1,{\bf k}_1} M^{[321]}_{{\bf k}_1,{\bf k}_1,{\bf 0}} \frac{n_{\rm B}[{\cal E}_{1,{\bf k}_1}] - n_{\rm B}[{\cal E}_{2,{\bf k}_1}]}{i\omega_n + {\cal E}_{2,{\bf k}_1} - {\cal E}_{1,{\bf k}_1}}, \nonumber \\
\Sigma^{(d)}_{11}(i\omega_n)&=- \frac{1}{N}\sum_{{\bf k}_1 \in \Lambda_0} M^{[331]}_{{\bf 0},{\bf 0},{\bf 0}} M^{[311]}_{{\bf k}_1,{\bf 0},{\bf k}_1} \frac{1}{\Delta} n_{\rm B}[{\cal E}_{1,{\bf k}_1}] - \frac{1}{N} \sum_{{\bf k}_1 \in \Lambda_0} M^{[311]}_{{\bf 0},{\bf 0},{\bf 0}} M^{[331]}_{{\bf 0},{\bf k}_1,{\bf k}_1} \frac{1}{\Delta} n_{\rm B}[{\cal E}_{1,{\bf k}_1}] \nonumber \\
&- \frac{1}{N}\sum_{{\bf k}_1 \in \Lambda_0} M^{[331]}_{{\bf 0},{\bf 0},{\bf 0}} M^{[412]}_{{\bf k}_1,{\bf 0},{\bf k}_1} \frac{1}{\Delta} n_{\rm B}[{\cal E}_{2,{\bf k}_1}] - \frac{1}{N} \sum_{{\bf k}_1 \in \Lambda_0} M^{[311]}_{{\bf 0},{\bf 0},{\bf 0}} M^{[432]}_{{\bf k}_1,{\bf 0},{\bf k}_1} \frac{1}{\Delta} n_{\rm B}[{\cal E}_{2,{\bf k}_1}]. \nonumber
\end{align}
\end{widetext}
Here the function $n_{\rm B}(x)=(e^{\beta x}-1)^{-1}$ is the Bose distribution function. We have defined a symmetrized third-order vertex coefficient
\begin{align}
M^{[l_1 l_2 l_3]}_{{\bf k}_1,{\bf k}_2,{\bf k}_3} 
=& M^{(l_1 l_2 l_3)}_{{\bf k}_1,{\bf k}_2,{\bf k}_3} + M^{(l_1 l_3 l_2)}_{{\bf k}_1,{\bf k}_3,{\bf k}_2} + M^{(l_2 l_1 l_3)}_{{\bf k}_2,{\bf k}_1,{\bf k}_3}  \nonumber \\
&+ M^{(l_2 l_3 l_1)}_{{\bf k}_2,{\bf k}_3,{\bf k}_1} + M^{(l_3 l_1 l_2)}_{{\bf k}_3,{\bf k}_1,{\bf k}_2} + M^{(l_3 l_2 l_1)}_{{\bf k}_3,{\bf k}_2,{\bf k}_1}.
\end{align}

Most dominant contributions to the decay of the Higgs mode stem from $\Sigma^{(b)}_{11}(i\omega_n)$. The partial function describes the Beliaev damping processes where one Higgs mode with zero momentum collapses into two NG modes with opposite momenta ${\bf k}$ and $-{\bf k}$ with satisfying the on-shell energy-momentum conservation of ${\cal E}_{1,{\bf 0}} - {\cal E}_{2,{\bf k}} - {\cal E}_{2,-{\bf k}}$. The Beliaev damping of the Higgs mode in the Bose-Hubbard systems has been studied in some literatures through calculations of its damping rate for $n_0 \gg 1$ at zero temperature \cite{Altman_2002} and at finite temperatures \cite{Nagao_2016}. 

In our previous study based on the imaginary-time Green's function theory \cite{Nagao_2016}, we have calculated the damping rate $\Gamma \equiv -{\rm Im}\Sigma_{11}(i\omega_{n})|_{i\omega_{n} \rightarrow \omega + i\epsilon}$ only at $\omega = {\cal E}_{1,{\bf 0}}=\Delta$ in order to obtain a qualitative measure of the stability of the Higgs mode. Our analysis of the present paper generalizes it such that the real and imaginary parts of the self-energy functions are taken into account at general frequencies. In particular, the real part is important because it characterizes a renormalization effect of the mean-field Higgs gap, which stems from interactions between the collective excitations.

The diagrams in Fig. \ref{fig: normal self-energy function} include processes with an odd number of NG modes. Such a contribution vanishes for a large filling rate $n_0 \gg 1$ due to the explicit particle-hole symmetry of the effective model ${\cal H}^{n_0 \gg 1}_{\rm eff}$. In particular, the contribution of $\Sigma^{(c)}_{11}$ can emerge only at $n_0 \sim 1$ and provides purely thermal effects on the damping properties of the Higgs mode. This process can be regarded as a Landau-type damping of the Higgs mode with absorbing one NG mode from a {\it thermal cloud} and emitting one Higgs mode. In Ref. \cite{Nagao_2016}, it has been reported that the NG mode with a non-zero momentum can exhibit a similar Landau damping into a single Higgs mode at finite temperatures via interactions with the NG modes in a thermal cloud. For basic explanations of the Landau damping of collective excitations, see, e.g., Ref.~\cite{Pethick_2008}.

\begin{figure*}
\includegraphics[width=150mm]{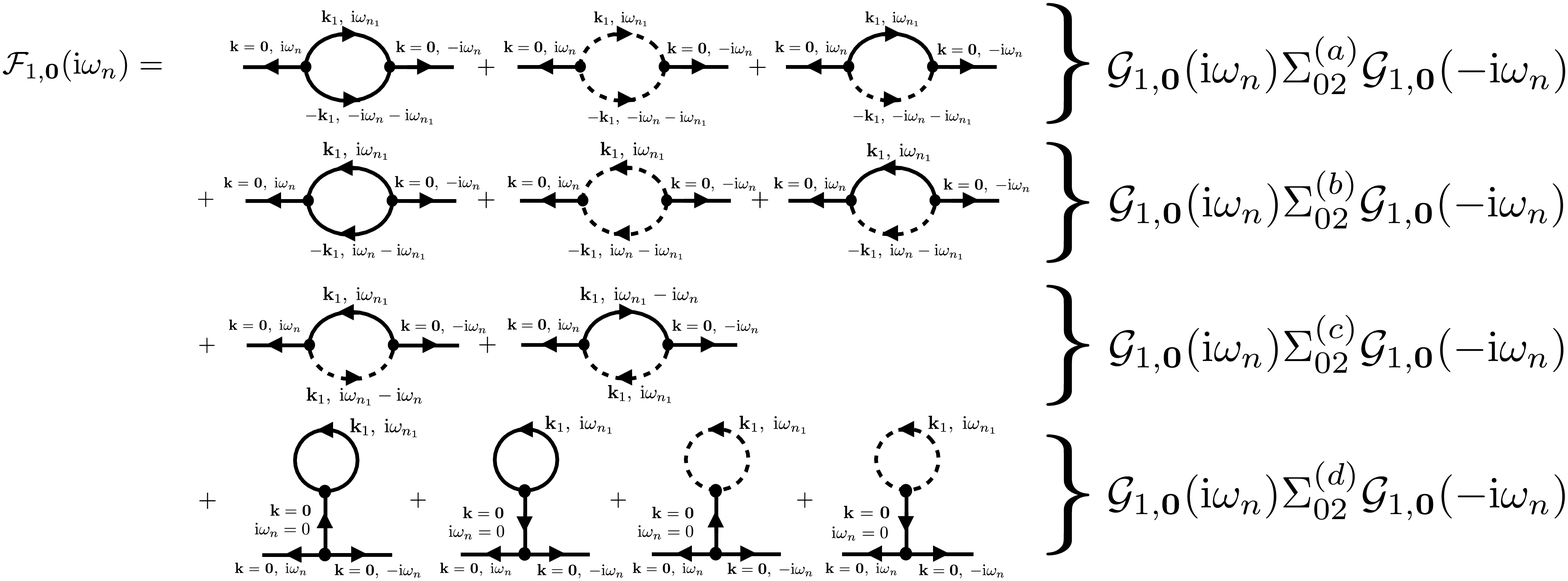}
\vspace{0mm}
\caption{Leading one-loop order contributions to the anomalous Green's function ${\cal F}_{1,{\bf 0}}(i\omega_n)$. The variables ${\bf k}_1$ and $i\omega_{n_1}$ added near the internal line implies the internal summation over the possible momentum and Matsubara frequency. The self-energy function of each diagram is obtained by picking off its two external lines. The diagrams in each column form an individual group labeled by $\Sigma^{(a)}_{02}(i\omega_n)$, $\Sigma^{(b)}_{02}(i\omega_n)$, $\Sigma^{(c)}_{02}(i\omega_n)$, or $\Sigma^{(d)}_{02}(i\omega_n)$.}
\label{fig: anomalous self-energy function}
\end{figure*}

We next calculate the anomalous self-energy function $\Sigma_{02}(i\omega_n)$. In Fig.~\ref{fig: anomalous self-energy function}, we depict the lowest order corrections to ${\cal F}_{1,{\bf 0}}(i\omega_n)$ by using the Feynman diagrams. The contributions to the anomalous self-energy function $\Sigma_{02}(i\omega_n)$ are also categorized into four groups as in the case of $\Sigma_{11}(i\omega_n)$; $\Sigma_{02}(i\omega_n) = \Sigma^{(a)}_{02}(i\omega_n) +  \Sigma^{(b)}_{02}(i\omega_n) +  \Sigma^{(c)}_{02}(i\omega_n) +  \Sigma^{(d)}_{02}(i\omega_n)$. Each diagram in $\Sigma_{02}(i\omega_n)$ has the same structure as the corresponding diagram in $\Sigma_{11}(i\omega_n)$ except for the interaction vertex on the righthand side at which the right external line connects with two internal lines. The analytic expressions of the anomalous self-energy function are given as follows:
\begin{widetext}
\begin{align}
\Sigma^{(a)}_{02}(i\omega_n) &=  -\frac{1}{2N} \sum_{{\bf k}_1 \in \Lambda_0} M^{[333]}_{{\bf 0},{\bf k}_1,-{\bf k}_1} M^{[311]}_{{\bf 0},{\bf k}_1,-{\bf k}_1} \frac{1+2n_{\rm B}[{\cal E}_{1,{\bf k}_1}]}{i\omega_n + 2{\cal E}_{1,{\bf k}_1} } -\frac{1}{2N} \sum_{{\bf k}_1 \in \Lambda_0} M^{[344]}_{{\bf 0},{\bf k}_1,-{\bf k}_1} M^{[322]}_{{\bf 0},{\bf k}_1,-{\bf k}_1} \frac{1+2n_{\rm B}[{\cal E}_{2,{\bf k}_1}]}{i\omega_n + 2{\cal E}_{2,{\bf k}_1} } \nonumber \\
& -\frac{1}{N} \sum_{{\bf k}_1 \in \Lambda_0} M^{[334]}_{{\bf 0},{\bf k}_1,-{\bf k}_1} M^{[312]}_{{\bf 0},{\bf k}_1,-{\bf k}_1} \frac{1+n_{\rm B}[{\cal E}_{1,{\bf k}_1}]+n_{\rm B}[{\cal E}_{2,{\bf k}_1}]}{i\omega_n + {\cal E}_{1,{\bf k}_1}+ {\cal E}_{2,{\bf k}_1} }, \nonumber \\  
\Sigma^{(b)}_{02}(i\omega_n) &= \frac{1}{2N} \sum_{{\bf k}_1 \in \Lambda_0} M^{[311]}_{{\bf 0},{\bf k}_1,-{\bf k}_1} M^{[333]}_{{\bf 0},{\bf k}_1,-{\bf k}_1} \frac{1+2n_{\rm B}[{\cal E}_{1,{\bf k}_1}]}{i\omega_n - 2{\cal E}_{1,{\bf k}_1} }   +\frac{1}{2N} \sum_{{\bf k}_1 \in \Lambda_0} M^{[322]}_{{\bf 0},{\bf k}_1,-{\bf k}_1} M^{[344]}_{{\bf 0},{\bf k}_1,-{\bf k}_1} \frac{1+2n_{\rm B}[{\cal E}_{2,{\bf k}_1}]}{i\omega_n - 2{\cal E}_{2,{\bf k}_1} }\nonumber \\
& +\frac{1}{N} \sum_{{\bf k}_1 \in \Lambda_0} M^{[312]}_{{\bf 0},{\bf k}_1,-{\bf k}_1} M^{[334]}_{{\bf 0},{\bf k}_1,-{\bf k}_1} \frac{1+n_{\rm B}[{\cal E}_{1,{\bf k}_1}]+n_{\rm B}[{\cal E}_{2,{\bf k}_1}]}{i\omega_n - {\cal E}_{1,{\bf k}_1} - {\cal E}_{2,{\bf k}_1} }, \nonumber \\
\Sigma^{(c)}_{02}(i\omega_n) &=  -\frac{1}{N} \sum_{{\bf k}_1 \in \Lambda_0} M^{[332]}_{{\bf 0},{\bf k}_1,{\bf k}_1} M^{[341]}_{{\bf 0},{\bf k}_1,{\bf k}_1} \frac{n_{\rm B}[{\cal E}_{1,{\bf k}_1}] - n_{\rm B}[{\cal E}_{2,{\bf k}_1}]}{i\omega_n + {\cal E}_{2,{\bf k}_1} - {\cal E}_{1,{\bf k}_1}} -\frac{1}{N} \sum_{{\bf k}_1 \in \Lambda_0} M^{[341]}_{{\bf 0},{\bf k}_1,{\bf k}_1} M^{[332]}_{{\bf 0},{\bf k}_1,{\bf k}_1} \frac{n_{\rm B}[{\cal E}_{2,{\bf k}_1}] - n_{\rm B}[{\cal E}_{1,{\bf k}_1}]}{i\omega_n + {\cal E}_{1,{\bf k}_1} - {\cal E}_{2,{\bf k}_1}}, \nonumber \\
\Sigma^{(d)}_{02}(i\omega_n) &= -\frac{1}{N}\sum_{{\bf k}_1 \in \Lambda_0} M^{[333]}_{{\bf 0},{\bf 0},{\bf 0}} M^{[311]}_{{\bf k}_1,{\bf k}_1,{\bf 0}} \frac{1}{\Delta} n_{\rm B}[{\cal E}_{1,{\bf k}_1}] -\frac{1}{N}\sum_{{\bf k}_1 \in \Lambda_0} M^{[331]}_{{\bf 0},{\bf 0},{\bf 0}} M^{[331]}_{{\bf 0},{\bf k}_1,{\bf k}_1} \frac{1}{\Delta} n_{\rm B}[{\cal E}_{1,{\bf k}_1}] \nonumber \\
&-\frac{1}{N}\sum_{{\bf k}_1 \in \Lambda_0} M^{[333]}_{{\bf 0},{\bf 0},{\bf 0}} M^{[421]}_{{\bf k}_1,{\bf k}_1,{\bf 0}} \frac{1}{\Delta} n_{\rm B}[{\cal E}_{2,{\bf k}_1}] -\frac{1}{N}\sum_{{\bf k}_1 \in \Lambda_0} M^{[331]}_{{\bf 0},{\bf 0},{\bf 0}} M^{[423]}_{{\bf k}_1,{\bf k}_1,{\bf 0}} \frac{1}{\Delta} n_{\rm B}[{\cal E}_{2,{\bf k}_1}]. \nonumber 
\end{align}
\end{widetext}

Here we should mention how to evaluate the momentum summations appearing in the self-energy functions. In our analysis, we have numerically computed the retarded self-energy functions such as $\Sigma^{\rm R}_{11}(\omega)=\Sigma_{11}(i\omega_n)|_{i\omega_n \rightarrow \omega + i\epsilon}$ for a fixed frequency after replacing the summations by the corresponding integral, i.e., $\sum_{{\bf k}_1 \in \Lambda_0} \rightarrow \int_{-\pi}^{\pi}\int_{-\pi}^{\pi}\int_{-\pi}^{\pi} dk_x dk_y dk_z/(2\pi)^3$. From the self-energy functions obtained numerically, we can construct the dynamical susceptibilities $\chi_{KK}(\omega)$ and $\chi_{OO}(\omega)$ according to the formulae (\ref{eq: tilde susceptibility_kk}) and (\ref{eq: tilde susceptibility_oo}).

\section{Results}\label{Sec: Results}

Using the formulations explained in the previous sections, we are able to obtain the dynamical susceptibilities. In this section, we compute the imaginary part of the susceptibility as a function of frequency $\omega$ and discuss the stability of the Higgs mode in three-dimensional optical lattice systems. 

\subsection{Response functions in the uniform system}\label{Sec: RFs_unif}

\begin{figure}
\includegraphics[width=80mm]{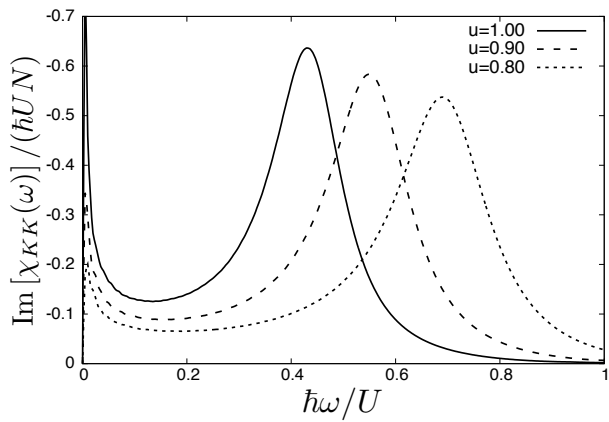}
\vspace{0mm}
\caption{Susceptibility at the unit filling rate $n_0=1$ and at zero temperature. The position of the resonance peak increases away from the critical point $u_c\approx 1.457$.}
\label{figure004}
\end{figure}

We analyze the response functions (dynamical susceptibilities) in the Bose-Hubbard model with no trapping potential in order to discuss the broadening of the resonance peak solely due to quantum and thermal fluctuations. In Fig. \ref{figure004}, we show ${\rm Im}\left[ \chi_{KK}(\omega)\right]$ at the unit filling rate and at zero temperature. Notice that the two response functions $\chi_{KK}(\omega)$ and $\chi_{OO}(\omega)$ are equal as mentioned in Sec.~\ref{Sec: LRA}. In Fig. \ref{figure004}, we see a sufficiently sharp resonance peak corresponding to the Higgs mode, which forms a Lorentzian-like curve. The center of the peak defines a renormalized Higgs gap and the width provides a damping rate of the mode. The existence of the sharp resonance peak implies that the Higgs mode is stable in the 3D system within the lowest order of the quantum fluctuation. In addition, the position of the peak shifts to the high-$\omega$ side as $u$ leaves from the critical point $u=u_{c}$. Table \ref{table001} shows each value of the renormalized Higgs gap $\Delta_{*}$ scaled by the corresponding Mott gap $\Delta_{\rm MI}$, which has a same relative distance from the critical point, ${\bar u}_{\rm rel} = |u-u_c|/u_c$, as that of the Higgs gap. As the energy scale, we used the mean-field Mott gap $\Delta_{\rm MI}=\sqrt{U^2-2JzU(2n_0+1)+(Jz)^2}$, which is derived in Ref. \cite{Huber_2007}.

\begin{figure}
\includegraphics[width=80mm]{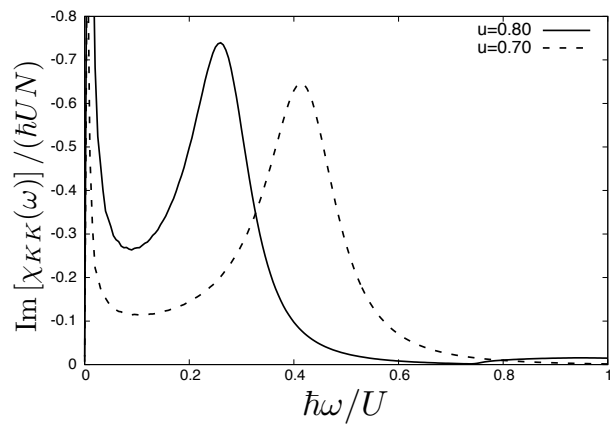}
\vspace{0mm}
\caption{Susceptibility at a large filling rate $n_0\gg1$ and at zero temperature. The peak position corresponding to the Higgs mode gap increases away from the critical point $u_c=1$.}
\label{figure005}
\end{figure}

We also see the similar behavior at a large filling rate $n_0\gg1$. In Fig.~\ref{figure005}, we show ${\rm Im}\left[ \chi_{KK}(\omega)\right]$ at a large filling rate and at zero temperature. The peak width approximately coincides with the damping rate of the Higgs mode evaluated in three dimensions and at a large filling rate \cite{Altman_2002}. It should be noted that we see another peak near $\omega = 0$. Such an additional peak near $\omega = 0$ also appears in Fig.~\ref{figure004}. The additional peak can be interpreted as an artifact of our perturbative method. In fact, the real parts of the self-energy functions become as large as the mean-field Higgs gap near $\omega = 0$. This means that the perturbative corrections to the Higgs gap are no longer small compared with the zeroth-order gap itself, i.e., the perturbative approximation breaks down near $\omega = 0$. Nevertheless, the perturbative corrections are sufficiently small compared to the Higgs gap $\Delta$ around $\omega = \Delta_{*}$, meaning that our perturbative approach is valid there. The emergence of this additional peak is related to the logarithmic infrared divergence of the self-energy functions of the $(3+1)$-dimensional relativistic $O(N)$ scalar model~\cite{Dupuis_2011}. Notice that in contrast to the infrared divergence of the self-energy function, our naive perturbation approach fails to describe the logarithmic corrections that appear as a result of renormalization of the marginal terms~\cite{Affleck_1992}, which is ignored in our analysis.

\begin{table}
\caption{The explicit values of the renormalized Higgs gap $\Delta_{*}$ scaled by the Mott gap $\Delta_{\rm MI}$. $\Delta_{*}$ and $\Delta_{\rm MI}$ locate at a same relative distance ${\bar u}_{\rm rel}$ from the critical point $u_c\approx 1.457$. $u_{o}$ and $u_{d}$ are the corresponding values of $u$ at a given ${\bar u}_{\rm rel}=|u-u_c|/u_c$ in the ordered side and disordered side, respectively.}
\begin{ruledtabular}
\begin{tabular}{ccccc}
$\Delta_{*}/\Delta_{\rm MI}$ & $\Delta/\Delta_{\rm MI}$ & ${\bar u}_{\rm rel}$ & $u_{o}$ & $u_{d}$ \\
\hline
0.890 & 1.081 & 0.314 & 1.000 & 1.914 \\
1.057 & 1.206 & 0.382 & 0.900 & 2.014  \\
1.251 & 1.359 & 0.451 & 0.800 & 2.114  \\
\end{tabular}
\end{ruledtabular}
\label{table001}
\end{table}

\begin{figure}
\includegraphics[width=80mm]{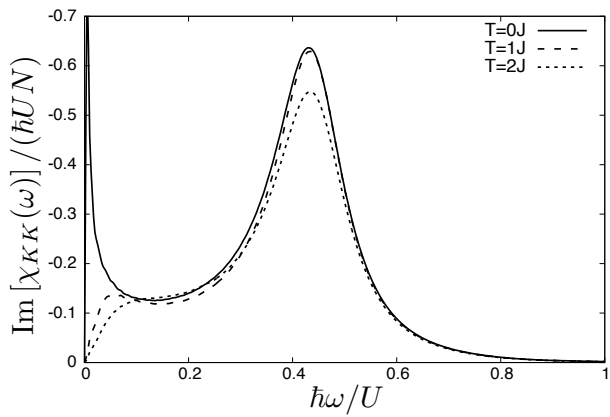}
\vspace{0mm}
\caption{Susceptibility at the unit filling rate $n_0=1$ and at typical temperatures. We have chosen the specific parameter $u=1$ ($zJ/U=0.25$).}
\label{fig: response function temperature}
\end{figure}

Next, we consider finite-temperature effects on the response functions. In Fig.~\ref{fig: response function temperature}, we show the temperature dependence of the susceptibility at $u=1$. The results show that the thermal fluctuation only makes the peak width slightly broader, so that the resonance peak is quite robust against the thermal fluctuations up to $T=2J$. Considering that the typical temperature in real experiments is of order $J$, we conclude that the Higgs resonance peak survives even at typical temperatures and at the unit filling rate. Our result is in contrast to the case of the 2D Bose-Hubbard model computed by the quantum Monte-Carlo simulations \cite{Pollet_2012}. 

It is worth noting that the damping rate of the zero-momentum Higgs mode at a large filling rate, which is computed in the similar way of our approaches used in this paper, also shows that the Higgs mode is sufficiently stable at typical temperatures of order $J$ \cite{Nagao_2016}. Our result presented in Fig. \ref{fig: response function temperature} generalizes the result obtained in the virtual large-filling case \cite{Nagao_2016} into a more realistic case with unit filling rate corresponding to actual experiments.

\subsection{Effects of a trapping potential}\label{Sec: LDA}

We include the trapping-potential effects within the local density approximation. As a specific shape of the potential, we assume a parabolic and isotropic potential
\begin{align}
V_{\rm trap}(r) = \frac{m\omega_0^2}{2}r^2, \label{eq: trap}
\end{align}
where $m$ is the atomic mass and $\omega_0$ is the frequency of the potential. According to the conventional local density approximation (LDA) \cite{Pitaevskii_2003}, the effect of the inhomogeneity is described by the general formula
\begin{align}
\chi_{\rm lda}(\omega) = 4\pi \int_{0}^{R} dr r^2 {\bar n}'[\mu(r)] \chi_{\rm unif}(\omega,\mu(r)), \label{eq: lda}
\end{align}
where $\chi_{\rm unif}(\omega,\mu)$ is the bulk susceptibility (Eqs. (\ref{eq: susceptibility_kk}) or (\ref{eq: susceptibility_oo}) divided by the factor $N$) at the fixed chemical potential $\mu$ and $\mu(r) = \mu_{n_0} - V_{\rm trap}(r)$ is the local chemical potential. ${\bar n}'[\mu]$ is the normalized density defined by 
\begin{align}
{\bar n}'[\mu(r)]=\frac{{\bar n}[\mu(r)]}{4\pi\int^{R}_{0} dr r^2 {\bar n}[\mu(r)]}.
\end{align}
$R$ denotes the radius of the spherical region, in which atoms are perturbed by the temporal modulation of $J$ or $U$. When the modulation perturbs the entire system, $R$ is equal to the Thomas--Fermi radius $R_{\rm TF}$, at which the density vanishes. We assume that at the trap center the density ${\bar n}[\mu_{n_0}]$ is tuned to $n_0=1$.

\subsubsection{Response functions at $R=R_{\rm TF}$}

We analyze the response function to the modulation applied globally to the entire system. In this case the radial integral in Eq.~(\ref{eq: lda}) is taken up to the Thomas--Fermi radius from the spatial center of the trapping potential: $R=R_{\rm TF}$.

\begin{figure*}
\includegraphics[width=140mm]{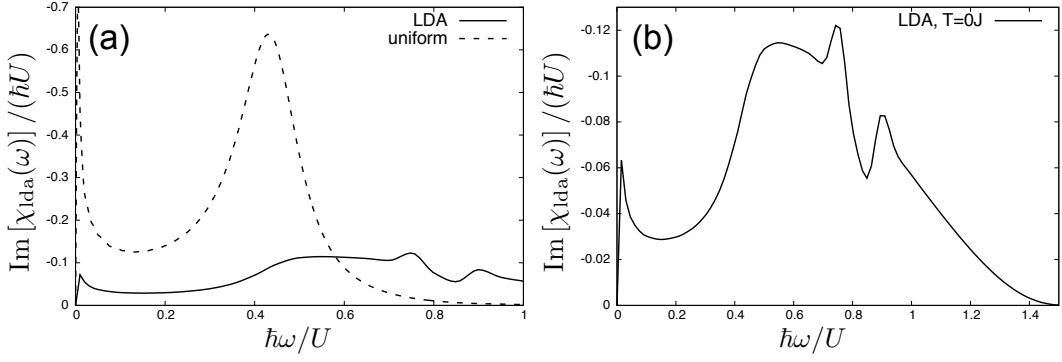}
\vspace{0mm}
\caption{(a) Dynamical susceptibility of the trapped system (the solid line) versus the one in the homogeneous system (the dashed line) at $T=0$ and $u=1$. At the center of the trap, the density of the system is tuned to unity, i.e., $n_0=1$. (b) Magnifying the dynamical susceptibility of the trapped system at zero temperature. In order to obtain a smooth line from LDA data, we used the spline interpolation. }
\label{figure007}
\end{figure*}

In Fig.~\ref{figure007} (a), we show ${\rm Im}\left[\chi_{\rm lda}(\omega)\right]$ together with the one in the absence of the potential (\ref{eq: trap}). We assume that $T=0$ and $u=1$, at which a sharp resonance peak survives when the system is homogeneous. In Fig.~\ref{figure007} (b), we plot the same LDA susceptibility in a magnified scale in order to see its detailed structure. There we see that the resonance peak, which would be rather sharp without the trapping potential, is significantly broadened due to the inhomogeneity effect so that the peak width is as large as the Higgs gap $\Delta$. In this sense, one can no longer regard the response as a well-defined resonance peak.

The broadening of the resonance peak can be attributed to the following reason. When we apply  the modulation globally to the entire system, all the subsystems corresponding to ${\bar n} \in [0,1]$ contribute to the resulting response. Specifically, the gap at $\bar{n}<1$ is larger than that at $\bar{n}=1$ and the high-energy contributions far from the trapping center obscure the well-defined Higgs resonance.

In Fig. \ref{figure007} (b), we also find a fine structure of the response in the region of $0.7U < \omega < 1.0 U$. This structure means that the response of the bulk gapful mode at a certain value of $\mu$, which gives $\Delta \simeq 0.75 U$ ($0.85 U$), is locally strong (weak). It is interesting to examine in future experiments whether or not the emergence of the fine structure is an artifact of LDA.

While the resonance peak structure in the response is smeared out, a characteristic feature of the Higgs mode in the bulk is still visible in the susceptibility of the trapped system. Specifically, the onset frequency of the response is almost equal to the bulk Higgs gap $\Delta$ at $\bar{n} = 1$. This property has been found also in 2D~\cite{Pollet_2012} and indeed utilized to measure $\Delta$ in experiment~\cite{Endres_2012}.
 
\subsubsection{Responses around the trapping center}

\begin{figure}
\includegraphics[width=80mm]{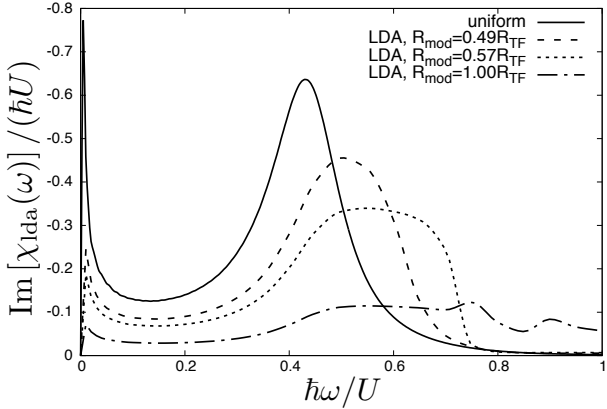}
\vspace{0mm}
\caption{Averaged dynamical susceptibilities in the trapped system modulated partially. The dashed, dotted, and dash-dotted lines correspond to $R_{\rm mod}/R_{\rm TF}= 0.49, 0.57, 1.00$, respectively. The susceptibility approaches the uniform result (solid line) in the limit of $R_{\rm mod}\rightarrow 0$. Here $u=1$ ($zJ/U=0.25$) and $T/J=0$. The filling factor at the trapping center is tuned to unity, i.e., $n_0=1$.}
\label{figure008}
\end{figure}

We analyze the response to a partial modulation, which acts only on atoms inside the spherical region with $R < R_{\rm TF}$ around the trap center. In this way, we eliminate the contributions from the low-density region that broaden the resonance peak and expect to see a sharp resonance peak as long as $R$ is sufficiently small. A similar analysis at 2D has been presented in Ref.~\cite{Liu_2015}. In what follows, we set $u=1$.

We define the radius for the partial modulation as $R_{\rm mod}$. In the unit of the Thomas-Fermi radius $R_{\rm TF}$, $R_{\rm mod}$ reads
\begin{align}
\frac{R_{\rm mod}}{R_{\rm TF}} = \sqrt{\frac{\mu_{n_0} - \mu_{\rm mod}}{\mu_{n_0} - \mu_{\rm TF}}},
\end{align}
where $\mu_{\rm mod} = \mu(R_{\rm mod})$ and $\mu_{\rm TF} = \mu(R_{\rm TF})$. In particular, one can easily see that $\mu_{\rm TF} = -0.75U$ for $u=1$. The calculation of the LDA is performed just by making $R=R_{\rm mod}$ in Eq.~(\ref{eq: lda}). 

\begin{figure}
\includegraphics[width=80mm]{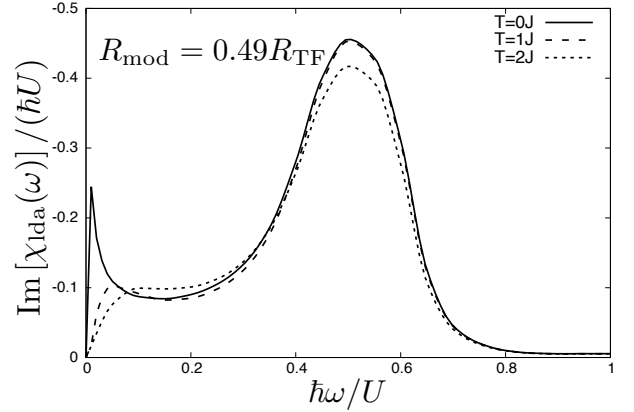}
\vspace{0mm}
\caption{Finite temperature effects on the susceptibilities of the trapped system at $u=1$ and $R_{\rm mod}/ R_{\rm TF} =0.49$. The solid, dashed, and dotted lines represent $T/J=0, 1, 2$, respectively. At the center of the trap, the density of the system is tuned to unity, i.e., $n_0=1$.}
\label{figure009}
\end{figure}

In Fig.~\ref{figure008}, we show ${\rm Im}\left[\chi_{\rm lda}(\omega)\right]$ at zero temperature for different values of $R_{\rm mod}$. When $R_{\rm mod}/R_{\rm TF}=0.49$, the modulation is added to a subregion of ${\bar n} \geq 0.90$. In this case, the shape of the resonance peak in the resulting response function is well approximated as a Lorentzian function and the peak width is clearly smaller than the peak position. Thus, the response exhibits a sharp resonance peak. We also find that the peak position is slightly shifted to the high-energy side due to the contribution from the low-density region.

When $R_{\rm mod}/R_{\rm TF}$ is increased, the response becomes broader to approach the result at $R_{\rm mod}=R_{\rm TF}$ (See the dash-dotted line in Fig.~\ref{figure008}). When $R_{\rm mod}/R_{\rm TF}=0.57$, at which a subregion of ${\bar n}(r) \geq 0.85$ is modulated, the response is significantly broader than that of $R_{\rm mod}/R_{\rm TF}=0.49$ and the shape of the response function noticeably deviates from a Lorentzian function. Thus, our results indicate that the condition that $R_{\rm mod}< 0.5 R_{\rm TF}$ is required for a sharp resonance peak to be observed. 

For the case of $R_{\rm mod}/R_{\rm TF}=0.49$, let us consider finite-temperature effects on the LDA susceptibility of the partial modulation. Figure \ref{figure009} shows the results at different temperatures. Our results reveal that the resonance peak is robust against thermal fluctuations at typical experimental temperatures of order $J$. The robustness of the response is not related with the specific choice of $R_{\rm mod}$ because the similar feature can be found in the uniform cases. According to the results in Fig. \ref{figure009}, it is expected to be able to detect the well-defined Higgs-mode resonance in typical three-dimensional experiments with a parabolic potential. The detection procedure requires a partial modulation of $J$ or $U$ over a radius $R_{\rm mod} \leq R_{\rm TF}$, and it is, in principle, possible in experiments. We emphasize that the temperature dependence in 3D systems is distinct from that in 2D systems \cite{Liu_2015}. In 2D systems, the response function significantly depends on the temperature so that the Higgs peak is smeared out due to thermal fluctuations when $T > J$ even for partial modulations.

\section{Conclusions}\label{Sec: conclusions}

In conclusion, we analyzed the effects of quantum and thermal fluctuations, and spatial inhomogeneity due to a trapping potential on the response functions of the 3D Bose-Hubbard model both for the hopping strength and onsite-interaction strength modulations, respectively. At the unit filling rate and in the absence of the trapping potential, our results showed that the Higgs mode can exist as a sharp resonance peak in the dynamical susceptibilities at typical temperatures. We included the effect of a trapping potential within the local density approximation and indicated that the resonance peak is significantly broadened due to the trapping potential when the modulations are applied globally to the entire system. In order to extract a sharp resonance peak from the smeared response, we discussed partial modulations around the trap center. The results with a modulation radius $R_{\rm mod}<0.5R_{\rm TF}$ showed that a well-defined resonance peak of the Higgs mode can survive at typical temperatures.

Recently, a quantum Monte-Carlo study on a 3D quantum antiferromagnet, which has a quantum critical point described effectively by the 3D relativistic $O(3)$ scalar model appears \cite{Qin_2017}. In this numerical work, some response functions are calculated and show a sufficiently sharp resonance peak of the Higgs mode. It is an interesting and important problem that one applies the same method to the 3D Bose-Hubbard model with the parabolic potential and test our qualitative results by utilizing such a more quantitative approach.

\begin{acknowledgments}
The authors thank D. Yamamoto and S. Nakajima for useful discussions. This work was supported by KAKENHI from Japan Society for the Promotion of Science: Grants No. 25220711 and CREST, JST No. JPMJCR1673.
\end{acknowledgments}

\appendix
\section{Energy absorption due to the onsite-interaction strength modulations} \label{App: onsite}

In this appendix we derive the relation between the response function (\ref{eq: response_oo}) and energy absorbed by the system for a finite-time period of the onsite-interaction strength modulation, according to the literatures about the hopping strength modulations \cite{Pekker_2015,Endres_thesis}.

As seen in Sec. \ref{Sec: modons}, the time-dependent Hamiltonian ${\cal H}_{\rm BH}(t) = {\cal H}_{\rm BH} + \Delta_{U}(t) O$ describes the behavior of the system that is initially in a thermal equilibrium state and is driven by the small and periodic modulation $U \rightarrow (1+\Delta_{U}(t))U = (1+\delta_{U}{\rm cos}(\omega t))U$ at a fixed $\omega$. If we assume that $\rho(t)$ is the total density operator at $t$, which approaches the equilibrium one $\rho_{\rm eq}$ as $t \rightarrow -\infty$, then the total energy of the system at $t$ is given by $E(t)=\langle {\cal H}_{\rm BH}(t) \rangle(t)={\rm Tr}\rho(t){\cal H}_{\rm BH}(t)$. We can verify easily that its instantaneous change rate $dE(t)/dt$ is proportional to only the instantaneous average of $O$ with a oscillation factor:
\begin{align}
\frac{dE}{dt} &= {\dot \Delta_U}(t) \langle O \rangle(t)  \nonumber \\
&= -\omega \delta_{U} {\rm sin}(\omega t) \langle O \rangle(t). \label{appendix: eq: dEdt}
\end{align}

Using the basic result of the linear response theory \cite{Altland_2010}, the response of $O$ to the $U$ modulation, it is defined by $\Delta\langle O \rangle(t) \equiv \langle O \rangle(t) -  \langle O \rangle_{\rm eq}$, is related to $\Delta_{U}(t)$ such as
\begin{align}
\Delta\langle O \rangle(t) = \int^{t}_{-\infty} D^{\rm R}_{OO}(t-t')\Delta_{U}(t'), \label{appendix: eq: deltaO}
\end{align}
where $D^{\rm R}_{OO}(t-t')$ is the response function given by Eq. (\ref{eq: response_oo}). Substituting $\Delta_{U}(t)=\delta_{U}{\rm cos}(\omega t)$ into this equation (\ref{appendix: eq: deltaO}), we obtain 
\begin{align}
\Delta\langle O \rangle(t) 
&= \delta_{U}{\rm Re}\left\{ e^{{\rm i}\omega t} \chi^*_{OO}(\omega) \right\} \nonumber \\
&= \delta_{U} \left\{ {\rm cos}(\omega t){\rm Re}\chi_{OO}(\omega) + {\rm sin}(\omega t){\rm Im}\chi_{OO}(\omega)  \right\}. \label{appendix: eq: deltaO_explicit}
\end{align}

Averaging Eq. (\ref{appendix: eq: dEdt}) over one period $t_{\rm mod} = 2\pi/\omega$ and using Eq. (\ref{appendix: eq: deltaO_explicit}), we finally obtain the mean energy absorbed by the system for a period of $t_{\rm mod}$
\begin{align}
\Delta E(\omega) = \frac{1}{t_{\rm mod}}\int^{t_{\rm mod}}_0 dt \frac{dE}{dt} = \frac{(\delta_{O})^2}{2} \omega S_{OO}(\omega), \label{appendix: eq: DeltaE}
\end{align}
where $S_{OO}(\omega) = -{\rm Im}\chi_{OO}(\omega)$ is the spectral function. One can measure $\Delta E(\omega)$ accurately by using the quantum-gas microscope technique. The relation (\ref{appendix: eq: DeltaE}) reveals that for the modulations of $U$, the experimental observable $\Delta E(\omega)$ is related only to the $O$-to-$O$ response function $D^{\rm R}_{OO}(t-t')$. 

\section{Supplement on the derivation of the effective pseudospin-one model}\label{App: effective model}

In the Hilbert space projected by ${\cal P}_{n_0}$, each of local operator that constitutes the model Hamiltonian, $a_{i}$, $a^{\dagger}_{i}$, and $\delta n = n_i - n_0$, reduces to a simple form represented by the constrained Schwinger bosons $t_{\alpha}$, $t^{\dagger}_{\alpha}$ ($\alpha = -1,0,1$). In terms of the bosons, the operators read
\begin{align}
{\cal P}_{n_0} a^{\dagger}_i {\cal P}_{n_0}^{-1}  &= \sqrt{n_0 + 1} t_{1,i}^{\dagger} t_{0,i} + \sqrt{n_0} t_{0,i}^{\dagger} t_{-1,i}, \nonumber \\
{\cal P}_{n_0} a_i {\cal P}_{n_0}^{-1} &= \sqrt{n_0 + 1} t_{0,i}^{\dagger} t_{1,i} + \sqrt{n_0} t_{-1,i}^{\dagger} t_{0,i},  \nonumber \\
{\cal P}_{n_0} \delta n_i {\cal P}^{-1}_{n_0} &= t_{1,i}^{\dagger}t_{1,i} - t_{-1,i}^{\dagger}t_{-1,i}. \nonumber 
\end{align}
For any nonzero and positive integer $n_0$, the last operator ${\cal P}_{n_0} \delta n_i {\cal P}^{-1}_{n_0}$ turns out to be the pseudospin-one operator $S^{z}_{i}$,
\begin{align}
{\cal P}_{n_0} \delta n_i {\cal P}^{-1}_{n_0} = S_i^z.
\end{align}
At $n_0 \gg 1$, due to $\sqrt{n_0 + 1} \approx \sqrt{n_0}$, we find that the remaining operators are rewritten by the pseudospin-one operators $S^{+}_{i}$ and $S^{-}_{i}$ simply, thus,
\begin{align}
{\cal P}_{n_0} a^\dagger_i {\cal P}^{-1}_{n_0} \approx \sqrt{\frac{n_0}{2}}S_i^+,\;\;\;{\cal P}_{n_0} a_i {\cal P}^{-1}_{n_0} \approx \sqrt{\frac{n_0}{2}}S_i^-. \label{Appendix: eq: a} 
\end{align}
Substituting these relations into Eq. (\ref{eq: def_eff}), we obtain the particle-hole symmetric effective pseudospin-one model (\ref{eq: high filling efm}).

On the other hand, for an arbitrary filling rate, the relations (\ref{Appendix: eq: a}) need to be modified. We can verify easily that $t^\dagger_{1,i}t_{0,i} = S^{z}_{i}S^{+}_{i}$ and $t^\dagger_{0,i}t_{1,i} = S^{-}_{i}S^{z}_{i}$, therefore, we obtain more complicated relations 
\begin{align}
{\cal P}_{n_0} a^\dagger_i {\cal P}^{-1}_{n_0} &= \sqrt{\frac{n_0}{2}}(1+\delta \nu S_i^z)S_i^+, \nonumber \\
{\cal P}_{n_0} a_i {\cal P}^{-1}_{n_0} &= \sqrt{\frac{n_0}{2}}S_i^-(1+\delta \nu S_i^z), 
\end{align}
where $\delta \nu = \sqrt{1 + 1/n_0}-1$. Substituting these relations into Eq. (\ref{eq: def_eff}), we obtain the explicit form of the modified effective pseudospin-one model (\ref{eq: low filling efm}). As seen in Sec. \ref{Sec: Methods}, this model has no longer the particle-hole symmetry when $\delta \nu \neq 0$.

\section{Supplement on the variational ansatz of the ground state wave function}\label{App: variational}

Using the variational wave function (\ref{eq: Gutzwiller}), the specific representation of the mean energy  density $E^{\rm MF} = \langle \Omega | {\cal H}^{n_0}_{\rm eff} |\Omega \rangle / N$ is computed as
\begin{align}
E^{\rm MF} = & \left[ \frac{1}{2} + \mu {\rm cos}\chi \right]{\rm sin}^2\left( \frac{\theta}{2} \right) \nonumber \\
&\;\;\;\;\; - \frac{Jz}{4}{\rm sin}^2\theta \left[ n_0+{\rm sin}^2\left( \frac{\chi}{2} \right) \right. \nonumber \\
&\;\;\;\;\;\;\;\;\;\;\;\;\;\;\; \left. + \sqrt{n_0(1+n_0)}{\rm sin}\chi {\rm cos}2\eta  \right]. \label{eq: energy}
\end{align}
After effecting the variation of Eq.~(\ref{eq: energy}) with respect to the variational parameters, we obtain a mean-field ground-state energy $E_0(\theta_{\rm mf}) = E^{\rm MF}(\theta_{\rm mf},0,0,\chi(\theta_{\rm mf}))$ where
\begin{align}
{\rm tan}\chi(\theta) = - \frac{2Jz\sqrt{n_0(n_0+1)}(1-{\rm sin}^2(\theta/2))}{2 \mu + Jz(1-{\rm sin}^2(\theta/2)) }, \label{eq: condition}
\end{align}
and $\theta_{\rm mf}$ is determined such that it minimizes the function $E_0(\theta)$. Using the optimized wave function after the variation, we also obtain the order parameter $\Psi = \langle \Omega | a_i | \Omega \rangle$ and mean density ${\bar n} = \langle \Omega | n_i | \Omega \rangle$ of the ground state as follows:
\begin{align}
\Psi = & \frac{1}{2}{\rm sin}\theta_{\rm mf}\left[ \sqrt{n_0+1} {\rm sin}\left(\frac{\chi_{\rm mf}}{2} \right) + \sqrt{n_0} {\rm cos}\left(\frac{\chi_{\rm mf}}{2} \right) \right], \nonumber \\ 
{\bar n} = & n_0 -{\rm sin}^2\left( \frac{\theta_{\rm mf}}{2} \right) {\rm cos}\chi_{\rm mf},\;\;\;\; \chi_{\rm mf}=\chi(\theta_{\rm mf}). \label{eq: density}
\end{align}

It is easy to obtain an analytical form of $\theta_{\rm mf}$ at commensurate filling rates. In this case, $\chi_{\rm mf}$ turns out to be $\chi_{\rm mf} = \pi/2$ (see Eq.~(\ref{eq: density})). Minimizing $E^{\rm MF}(\theta,0,0,\pi/2)$ with respect to $\theta$, we obtain 
\begin{align}
\theta_{\rm mf} = {\rm sin}^{-1}\left( \sqrt{1 - (Jz)^{-2}(\sqrt{n_0 + 1} + \sqrt{n_0})^{-4}} \right), 
\end{align}
and the corresponding chemical potential at ${\bar n} = n_0$ reads
\begin{align}
\mu_{n_0}=-\frac{1}{4}\left[ zJ+(\sqrt{n_0+1}+\sqrt{n_0})^{-2} \right].
\end{align}

Here, it is worth noting that at $\chi_{\rm mf} = \pi/2$ the ground state is particle-hole symmetric. This is because the corresponding wave function (\ref{eq: Gutzwiller}) contains $t^\dagger_{1,i}$ and $t^\dagger_{-1,i}$ components with equal weights at each site.

\section{Coefficients in the effective model}\label{App: coefficients}
In this appendix we give the coefficients in each partial Hamiltonian ${\cal H}^{(l)}_{\rm eff}$ for $l=0,1,2,3,4$. To simplify our discussion, we define a formal representation of the pseudospin operators as follows:
\begin{align}
S^{+}_{i}&=t^{\dagger}_{i}T_{1}t_{i},\;\;S^{-}_{i}=t^{\dagger}_{i}T_{2}t_{i},\;\;S^{z}_{i}=t^{\dagger}_{i}T_{3}t_{i}, \nonumber \\
(S^{z}_{i})^2&=t^{\dagger}_{i}T_{4}t_{i},\;\;S^{z}_{i}S^{+}_{i} = t^{\dagger}_{i}T_{5}t_{i},\;\;S^{-}_{i}S^{z}_{i} = t^{\dagger}_{i}T_{6}t_{i},
\end{align}
where $t_{i}=(t_{1,i},t_{0,i},t_{-1,i})^{\rm T}$. We have introduced matrices $T_{1},T_{2},\cdots,T_{6}$ defined by 
\begin{align}
T_1 &=
\begin{pmatrix}
0 & \sqrt{2} & 0 \\
0 &  0 & \sqrt{2} \\
0 & 0 & 0
\end{pmatrix}
,\;\;
T_2 =
\begin{pmatrix}
0 &  0 & 0 \\
\sqrt{2} &  0 & 0 \\
0 & \sqrt{2} & 0
\end{pmatrix}
,\nonumber \\
T_3 &=
\begin{pmatrix}
1 &  0 &  0 \\
0 &  0 &  0 \\
0 &  0 & -1
\end{pmatrix}
,\;\;
T_4 =
\begin{pmatrix}
1 &  0 &  0 \\
0 &  0 &  0 \\
0 &  0 & 1
\end{pmatrix}
,\nonumber \\
T_5 &=
\begin{pmatrix}
0 & \sqrt{2} & 0 \\
0 &         0  & 0 \\
0 &         0  & 0
\end{pmatrix}
,\;\;
T_6 =
\begin{pmatrix}
           0 & 0 & 0 \\
 \sqrt{2} & 0  & 0 \\
          0 & 0  & 0
\end{pmatrix}
.
\end{align}

The canonical transformation (\ref{eq: canonical}) can be regarded as the linear transformation from the old basis $t_{i}$ to the new one $b_{i}=(b_{1,i},b_{0,i},b_{2,i})^{\rm T}$. After the transformation, the elements of the matrices in the new basis are given by 
\begin{align}
{\tilde T}_1 &=
\begin{pmatrix}
-\sqrt{2}s_{1}c_{1}(s_{2}+c_{2})              & \sqrt{2}({s_{1}}^2 c_{2} - {c_{1}}^2 s_{2}) & -\sqrt{2}s_{1}s_{2} \\
 \sqrt{2}({s_{1}}^2 s_{2} - {c_1}^2 c_{2}) &                 \sqrt{2}s_{1}c_{1}(s_{2}+c_{2})& -\sqrt{2}c_{1}s_{2} \\
\sqrt{2}s_{1}c_{2}                                    &                                    \sqrt{2}c_{1}c_{2} &                          0 
\end{pmatrix}
,\nonumber \\
{\tilde T}_2 &=
\begin{pmatrix}
-\sqrt{2}s_{1}c_{1}(s_{2}+c_{2})              & \sqrt{2}({s_{1}}^2 s_{2} - {c_{1}}^2 c_{2}) & \sqrt{2}s_{1}c_{2}  \\
 \sqrt{2}({s_1}^2 c_{2} - {c_{1}}^2 s_{2}) &         \sqrt{2}s_{1}c_{1}(s_{2}+c_{2})        &  \sqrt{2}c_{1}c_{2}  \\
-\sqrt{2}s_{1}s_{2}                                  &                           -\sqrt{2}c_{1}s_{2}          &                          0
\end{pmatrix}
,\nonumber \\
{\tilde T}_3 &=
\begin{pmatrix}
{c_{1}}^2({s_{2}}^2 - {c_{2}}^2)    & s_{1}c_{1}({c_{2}}^2 - {s_{2}}^2)  & -2c_{1}s_{2}c_{2}\\
s_{1}c_{1}({c_{2}}^2 - {s_{2}}^2)  & {s_{1}}^2({s_{2}}^2 - {c_{2}}^2) & 2s_{1}s_{2}c_{2}  \\
-2c_{1}s_{2}c_{2}                        & 2s_{1}s_{2}c_{2}                & {c_{2}}^2 - {s_{2}}^2
\end{pmatrix}
,\nonumber \\
{\tilde T}_4 &=
\begin{pmatrix}
{c_{1}}^2     & -s_{1}c_{1}   & 0 \\
-s_{1}c_{1} &   {s_{1}}^2      & 0 \\
0                 &     0               & 1 
\end{pmatrix}
,\nonumber \\
{\tilde T}_5 &=
\begin{pmatrix}
-\sqrt{2}s_{1}c_{1}s_{2} & -\sqrt{2}{c_{1}}^2s_{2} & 0 \\
\sqrt{2}{s_{1}}^2s_{2} & \sqrt{2}s_{1}c_{1}s_{2}     & 0 \\
\sqrt{2}s_{1}c_{2}    & \sqrt{2}c_{1}c_{2}      & 0
\end{pmatrix}
,\nonumber \\
{\tilde T}_6 &=
\begin{pmatrix}
-\sqrt{2}s_{1}c_{1}s_{2}    & \sqrt{2}{s_{1}}^2 s_{2}& \sqrt{2}s_{1}c_{2} \\
-\sqrt{2}{c_{1}}^2 s_{2} &  \sqrt{2}s_{1}c_{1}s_{2}  & \sqrt{2}c_{1}c_{2} \\
                             0  &                           0   &                       0
\end{pmatrix}
.
\end{align}
In the following equations, we express the matrix elements of each matrix by 
\begin{align}
{\tilde T}_{\mu} &=
\begin{pmatrix}
({\tilde T}_{\mu})_{11} & ({\tilde T}_{\mu})_{10} & ({\tilde T}_{\mu})_{12} \\
({\tilde T}_{\mu})_{01} & ({\tilde T}_{\mu})_{00} & ({\tilde T}_{\mu})_{02} \\
({\tilde T}_{\mu})_{21} & ({\tilde T}_{\mu})_{20} & ({\tilde T}_{\mu})_{22} 
\end{pmatrix}
,\;{\rm for}\;\mu=1,2,\cdots,6. \nonumber
\end{align}

In terms of the matrix elements, the coefficients in ${\cal H}_{\rm eff}^{(0)}$ are given by 
\begin{align}
A_0&=-\frac{Jn_0z}{2}\{({\tilde T}_1)_{00} + \delta \nu ({\tilde T}_5)_{00}\}^2, \\
{\tilde A}_0 &= \frac{U}{2}({\tilde T}_4)_{00} - B({\tilde T}_3)_{00}. 
\end{align}
The coefficients in ${\cal H}_{\rm eff}^{(1)}$ are given by 
\begin{align}
A_1&= -\frac{Jn_0z}{2} \{({\tilde T}_1)_{00} + \delta \nu ({\tilde T}_5)_{00}\} \nonumber \\
&\;\;\;\;\; \times \left[ ({\tilde T}_1)_{01} + ({\tilde T}_1)_{10} + \delta \nu ({\tilde T}_5)_{01}  + \delta \nu ({\tilde T}_5)_{10} \right], \\
B_1&= -\frac{Jn_0z}{2} \{({\tilde T}_1)_{00} + \delta \nu ({\tilde T}_5)_{00}\} \nonumber \\
&\;\;\;\;\; \times \left[ ({\tilde T}_1)_{02} + ({\tilde T}_1)_{20} + \delta \nu ({\tilde T}_5)_{20} \right],  \\
{\tilde A}_1 &= \frac{U}{2}({\tilde T}_4)_{10} - B ({\tilde T}_3)_{10}, \\
{\tilde B}_1 &= - B ({\tilde T}_3)_{20}.
\end{align}
The coefficients in ${\cal H}_{\rm eff}^{(2)}$ are given by 
\begin{align}
A_2 &= -Jn_0z\{({\tilde T}_1)_{00} + \delta \nu ({\tilde T}_5)_{00}\} \nonumber \\
&\;\;\;\;\;\;\;\;\;\;\;\;\;\;\;\;\;\;\;\;\;\;\;\;\; \times \{({\tilde T}_1)_{11} + \delta \nu ({\tilde T}_5)_{11}\}, \\
B_2 &= -\frac{Jn_0z}{2}\left[ \{({\tilde T}_1)_{00} + \delta \nu ({\tilde T}_5)_{00}\}\{({\tilde T}_1)_{21} + \delta \nu ({\tilde T}_5)_{21}\} \right. \nonumber \\
&\;\;\;\;\;\;\;\;\;\;\;\;\;\;\;\;\;\;\;\;\;\; + \left.({\tilde T}_1)_{12}\{({\tilde T}_1)_{00} + \delta \nu ({\tilde T}_5)_{00}\} \right], \\
D_2 &= -\frac{Jn_0z}{2}\{({\tilde T}_1)_{10} + \delta \nu ({\tilde T}_5)_{10}\}  \nonumber \\
&\;\;\;\;\;\;\;\;\;\;\;\;\;\;\;\;\;\;\;\;\;\;\;\;\; \times\{({\tilde T}_1)_{01} + \delta \nu ({\tilde T}_5)_{01}\}, \\
E_2 &= -\frac{Jn_0z}{2} \left[ \{ ({\tilde T}_1)_{10} + \delta \nu ({\tilde T}_5)_{10}\}^2 \right. \nonumber \\
&\;\;\;\;\;\;\;\;\;\;\;\;\;\;\;\;\;\;\;\;\;\;\;\;\;+ \left. \{ ({\tilde T}_1)_{01} + \delta \nu ({\tilde T}_5)_{01} \}^2 \right], \\
F_2 &= -\frac{Jn_0z}{2} \left[ \{({\tilde T}_1)_{20} + \delta \nu ({\tilde T}_5)_{20}\}\{({\tilde T}_1)_{10} + \delta \nu ({\tilde T}_5)_{10}\} \right. \nonumber \\
&\;\;\;\;\;\;\;\;\;\;\;\;\;\;\;\;\;\;\;\;+ \left. ({\tilde T}_1)_{02}\{({\tilde T}_1)_{01} + \delta \nu ({\tilde T}_5)_{01}\} \right], \\
G_2 &= -\frac{Jn_0z}{2} \left[ \{({\tilde T}_1)_{20} + \delta \nu ({\tilde T}_5)_{20}\}\{({\tilde T}_1)_{01} + \delta \nu ({\tilde T}_5)_{01}\} \right. \nonumber \\
&\;\;\;\;\;\;\;\;\;\;\;\;\;\;\;\;\;\;\;\;+ \left. ({\tilde T}_1)_{02}\{({\tilde T}_1)_{10} + \delta \nu ({\tilde T}_5)_{10}\}\right], \\
H_2 &= -\frac{Jn_0z}{2} ({\tilde T}_1)_{02}\{({\tilde T}_1)_{20} + \delta \nu ({\tilde T}_5)_{20}\}, \\
I_2 &= -\frac{Jn_0z}{2} \left[ \{({\tilde T}_1)_{20} + \delta \nu ({\tilde T}_5)_{20}\}^2 + ({\tilde T}_1)_{02}^2\right], \\
{\tilde A}_2 &= \frac{U}{2}({\tilde T}_4)_{11} - B({\tilde T}_3)_{11}, \\
{\tilde B}_2 &= -B({\tilde T}_3)_{12}, \\
{\tilde C}_2 &= \frac{U}{2}({\tilde T}_4)_{22} - B({\tilde T}_3)_{22}.
\end{align}
The coefficients in ${\cal H}_{\rm eff}^{(3)}$ are given by 
\begin{align}
A_3 &= -\frac{Jn_0z}{2}\{({\tilde T}_1)_{11}+\delta\nu ({\tilde T}_5)_{11}\} \nonumber \\
&\;\;\; \times\{({\tilde T}_1)_{10} + ({\tilde T}_1)_{01} + \delta\nu ({\tilde T}_5)_{10}+\delta\nu ({\tilde T}_5)_{01}\}, \\
B_3 &= -\frac{Jn_0z}{2}\{({\tilde T}_1)_{11}+\delta\nu ({\tilde T}_5)_{11}\} \nonumber \\
&\;\;\;\;\;\;\;\;\;\;\;\;\; \times\{({\tilde T}_1)_{20} + ({\tilde T}_1)_{02} + \delta\nu ({\tilde T}_5)_{20}\}, \\
C_3 &= -\frac{Jn_0z}{2} \left[ \{({\tilde T}_1)_{10}+\delta\nu ({\tilde T}_5)_{10})({\tilde T}_1)_{12} \right. \nonumber \\
&\; + \left. \{({\tilde T}_1)_{21}+\delta\nu ({\tilde T}_5)_{21}\}\{({\tilde T}_1)_{01}+\delta\nu ({\tilde T}_5)_{01}\} \right], \\
D_3 &= -\frac{Jn_0z}{2} \left[ \{({\tilde T}_1)_{10}+\delta\nu ({\tilde T}_5)_{10}\}\{({\tilde T}_1)_{21}+\delta\nu ({\tilde T}_5)_{21}\} \right. \nonumber \\
&\;\;\;  + \left. \{{\tilde T}_1)_{12}(({\tilde T}_1)_{01}+\delta\nu ({\tilde T}_5)_{01}\} \right], \\
E_3 &= -\frac{Jn_0z}{2} \left[\{({\tilde T}_1)_{21}+\delta\nu ({\tilde T}_5)_{21}\}\{({\tilde T}_1)_{20} + \delta\nu ({\tilde T}_5)_{20}\} \right. \nonumber \\
&\;\;\;\;\;\;\;\;\;\;\;  + \left. ({\tilde T}_1)_{02}({\tilde T}_1)_{12} \right], \\
F_3 &= -\frac{Jn_0z}{2} \left[ ({\tilde T}_1)_{02}\{({\tilde T}_1)_{21} + \delta\nu ({\tilde T}_5)_{21}\} \right. \nonumber \\
&\;\;\;\;\;\;\;\;\;\;\; + \left. ({\tilde T}_1)_{12}\{({\tilde T}_1)_{20} + \delta\nu ({\tilde T}_5)_{20}\} \right],
\end{align}
Finally, the coefficients in ${\cal H}_{\rm eff}^{(4)}$ are given by 
\begin{align}
A_4 &= -\frac{Jn_0z}{2}\{({\tilde T}_1)_{11} + \delta \nu ({\tilde T}_5)_{11}\}^2,  \\
B_4 &= -\frac{Jn_0z}{2}({\tilde T}_1)_{12}\{({\tilde T}_1)_{21} + \delta \nu ({\tilde T}_5)_{21}\},  \\
C_4 &= -\frac{Jn_0z}{2} \{({\tilde T}_1)_{11} + \delta \nu ({\tilde T}_5)_{11}\} \nonumber \\
&\;\;\;\;\;\;\;\;\;\;\; \times \left[ ({\tilde T}_1)_{12} + ({\tilde T}_1)_{21} + \delta \nu ({\tilde T}_5)_{21} \right],   \\
D_4 &= -\frac{Jn_0z}{2} \left[  \{({\tilde T}_1)_{21} + \delta \nu ({\tilde T}_5)_{21}\}^2 +  ({\tilde T}_1)_{12}^2 \right].
\end{align}

\section{Bogoliubov transformation at large filling rates} \label{App: bogoliubov}

At $n_0 \gg 1$, we can compute ${\rm W}_{\bf k}$, ${\cal E}_{1, \bf k}$,  and ${\cal E}_{2, \bf k}$ analytically. As we have seen in Sec. \ref{Sec: HPE}, ${\cal H}^{(2)}_{\rm SW}$ has no mixing term between branches labeled by 1 or 2 in the limit. Hence, we can perform the Bogoliubov transformation independently in each blanch:
\begin{align}
{\rm W}_{\bf k}
\rightarrow
\begin{pmatrix}
u_{11,{\bf k}}  & 0                      & v^{*}_{11,-{\bf k}} & 0 \\
0                   & u_{22,{\bf k}}     & 0                         & v^{*}_{22,-{\bf k}}  \\
v_{11,{\bf k}}   & 0                     & u^*_{11,-{\bf k}}    & 0 \\
0                   & v_{22,{\bf k}}     & 0                         & u^*_{22,-{\bf k}} 
\end{pmatrix}
. \nonumber
\end{align}

Let us assume that the coefficients are real and have a symmetry under a sign change of the momentum ${\bf k}\rightarrow -{\bf k}$. In this assumption, the coefficients of the transformation are
\begin{align}
u_{11,{\bf k}}&=\sqrt{\frac{2-u^2\gamma_{\bf k}}{4\sqrt{1-u^2\gamma_{\bf k}}}+\frac{1}{2}},\\
v_{11,{\bf k}}&={\rm sgn}({\gamma_{\bf k}})\sqrt{\frac{2-u^2\gamma_{\bf k}}{4\sqrt{1-u^2\gamma_{\bf k}}}-\frac{1}{2}},\\
u_{22,{\bf k}}&=\sqrt{\frac{2-\gamma_{\bf k}}{4\sqrt{1-\gamma_{\bf k}}}+\frac{1}{2}},\\
v_{22,{\bf k}}&=-{\rm sgn}({\gamma_{\bf k}})\sqrt{\frac{2-\gamma_{\bf k}}{4\sqrt{1-\gamma_{\bf k}}}-\frac{1}{2}}.
\end{align}
The band dispersions of the Higgs and NG modes in the large filling limit \cite{Altman_2002} are
\begin{align}
{\cal E}_{1,{\bf k}}&= 2Jn_0z\sqrt{1-u^2 \gamma_{\bf k}}, \\
{\cal E}_{2,{\bf k}}&= Jn_0z(1+u)\sqrt{1 - \gamma_{\bf k}}.
\end{align}
The former Higgs band has a finite energy gap ${\tilde \Delta}=2Jn_0z\sqrt{1-u^2}$ at ${\bf k}=0$ while the latter NG band is gapless. The energy gap ${\tilde \Delta}$ closes at the critical point $u=u_{c}=1$.

\newpage 
\bibliography{apssamp}

\begin{thebibliography}{99}
\bibitem{Volovik_2014} G. E.Volovik and M. A. Zubkov, J. Low Temp. Phys. {\bf 175}, 486 (2014).
\bibitem{Pekker_2015} D. Pekker and C. M. Varma, Annu. Rev. Condens. Matter Phys. {\bf 6}, 269 (2015).
\bibitem{Higgs_1964} P. W. Higgs, Phys. Rev. Lett. {\bf 13}, 508 (1964).
\bibitem{Sooryakumar_1980} R. Sooryakumar and M. V. Klein, Phys. Rev. Lett. {\bf 45}, 660 (1980). 
\bibitem{Sooryakumar_1981} R. Sooryakumar and M. V. Klein, Phys. Rev. B {\bf 23}, 3213 (1981). 
\bibitem{Littlewood_1981} P. B. Littlewood and C. M. Varma, Phys. Rev. Lett. {\bf 47}, 811 (1981). 
\bibitem{Littlewood_1982} P. B. Littlewood and C. M. Varma, Phys. Rev. B {\bf 26}, 4883 (1982).
\bibitem{Measson_2014} M.-A. M\'{e}asson, Y. Gallais, M. Cazayous, B. Clair, P. Rodi\'{e}re, L. Cario, and A. Sacuto, Phys. Rev. B {\bf 89}, 060503 (2014).
\bibitem{Matsunaga_2013} R. Matsunaga, Y. I. Hamada, K. Makise, Y. Uzawa, H. Terai, Z. Wang, and R. Shimano, Phys. Rev. Lett. {\bf 111}, 057002 (2013).
\bibitem{Matsunaga_2014} R. Matsunaga, N. Tsuji, H. Fujita, A. Sugioka, K. Makise, Y. Uzawa, H. Terai, Z. Wang, H. Aoki, and R. Shimano, Science {\bf 345}, 6201 (2014).
\bibitem{Sherman_2015} D. Sherman, U. S. Pracht, B. Gorshunov, S. Poran, J. Jesudasan, M. Chand, P. Raychaudhuri, M. Swanson, N. Trivedi, A. Auerbach, M. Scheffler, A. Frydman, and M. Dressel, Nat. Phys. {\bf 11}, 188 (2015).
\bibitem{Matsunaga_2017} R. Matsunaga, N. Tsuji, K. Makise, H. Terai, H. Aoki, and R. Shimano, Phys. Rev. B {\bf 96}, 020505 (2017).
\bibitem{Ruegg_2008} Ch. R\"{u}egg, B. Normand, M. Matsumoto, A. Furrer, D. F. McMorrow, K.W. Kramer, H. U. Gudel, S. N. Gvasaliya, H. Mutka, and M. Boehm, Phys. Rev. Lett. {\bf 100}, 205701 (2008).
\bibitem{Merchant_2014} P. Merchant, B. Normand, K. W. Kr\"{a}mer, M. Boehm, D. F. McMorrow, and Ch. R\"{u}egg, Nat. Phys. {\bf10}, 373 (2014).
\bibitem{Kuroe_2014} H. Kuroe, N. Takami, N. Niwa, T. Sekine, M. Matsumoto, F. Yamada, H. Tanaka, and K. Takemura, J. Phys.: Conf. Series {\bf 400}, 032042 (2012).
\bibitem{Demsar_1999} J. Demsar, K. Biljakovi\'{c}, and D. Mihailovic, Phys. Rev. Lett. {\bf 83}, 800 (1999).
\bibitem{Schaefer_2014} H. Schaefer, V. V. Kabanov, and J. Demsar, Phys. Rev. B {\bf 89}, 045106 (2014).
\bibitem{Yusupov_2010} R. Yusupov, T. Mertelj, V. V. Kabanov, S. Brazovskii, P. Kusar, J.-H. Chu, I. R. Fisher, and D. Mihailovic, Nat. Phys. {\bf 6}, 681 (2010).
\bibitem{Mertelj_2013} T. Mertelj, P. Kusar, V. V. Kabanov, P. Giraldo-Gallo, I. R. Fisher, and D. Mihailovic, Phys. Rev. Lett. {\bf 110}, 156401 (2013).
\bibitem{Avenel_1980} O. Avenel, E. Varoquaux, and H. Ebisawa, Phys. Rev. Lett. {\bf 45}, 1952 (1980).
\bibitem{Collett_2013} C. A. Collett, J. Pollanen, J. I. A. Li, W. J. Gannon, and W. P. Halperin, J. Low Temp. Phys. {\bf 171}, 214 (2013).
\bibitem{Bissbort_2011} U. Bissbort, S. G\"{o}tze, Y. Li, J. Heinze, J. S. Krauser, M. Weinberg, C. Becker, K. Sengstock, and W. Hofstetter, Phys. Rev. Lett. {\bf 106}, 205303 (2011).
\bibitem{Endres_2012} M. Endres, T. Fukuhara, D. Pekker, M. Cheneau, P. Schau\ss, C. Gross, E. Demler, S. Kuhr, and I. Bloch, Nature {\bf 487}, 454 (2012).
\bibitem{Podolsky_2011} D. Podolsky, A. Auerbach, and D. P. Arovas, Phys. Rev. B {\bf 84}, 174522 (2011).
\bibitem{Podolsky_2012} D. Podolsky and S. Sachdev, Phys. Rev. B {\bf 86}, 054508 (2012).
\bibitem{Gazit_2013} S. Gazit, D. Podolsky, and A. Auerbach, Phys. Rev. Lett. {\bf 110}, 140401 (2013); S. Gazit, D. Podolsky, A. Auerbach, and D. P. Arovas, Phys. Rev. B {\bf 88}, 235108 (2013).
\bibitem{Rancon_2014} A. Ran\c{c}on and N. Dupuis, Phys. Rev. B {\bf 89}, 180501(R) (2014).
\bibitem{Katan_2015} Y. T. Katan and D. Podolsky, Phys. Rev. B {\bf 91}, 075132 (2015).
\bibitem{Rose_2015} F. Rose, F. L\'{e}onard, and N. Dupuis, Phys. Rev. B {\bf 91}, 224501 (2015).
\bibitem{Pollet_2012} L. Pollet and N. Prokof’ev, Phys. Rev. Lett. {\bf 109}, 010401 (2012).
\bibitem{Liu_2015} L. Liu, K. Chen, Y. Deng, M. Endres, L. Pollet, and N. Prokof'ev, Phys. Rev. B {\bf 92}, 174521 (2015).
\bibitem{Chen_2013} K. Chen, L. Liu, Y. Deng, L. Pollet, and N. Prokof’ev, Phys. Rev. Lett. {\bf 110}, 170403 (2013).
\bibitem{Altman_2002} E. Altman and A. Auerbach, Phys. Rev. Lett. {\bf 89}, 250404 (2002).
\bibitem{Nagao_2016} K. Nagao and I. Danshita, Prog. Theor. Exp. Phys. {\bf 2016}, 063I01 (2016).
\bibitem{Huber_2007} S. D. Huber, E. Altman, H. P. B\"{u}chler, and G. Blatter, Phys. Rev. B {\bf 75}, 085106 (2007).
\bibitem{Fisher_1989} M. P. A. Fisher, P. B. Weichman, G. Grinstein, and D. S. Fisher, Phys. Rev. B {\bf 40}, 546 (1989).
\bibitem{Jaksch_1998} D. Jaksch, C. Bruder, J. I. Cirac, C. W. Gardiner, and P. Zoller, Phys. Rev. Lett. {\bf 81}, 3108 (1998).
\bibitem{Oosten_2001} D. van Oosten, P. van der Straten, and H. T. C. Stoof, Phys. Rev. A {\bf 63}, 053601 (2001).
\bibitem{Sansone_2007} B. Capogrosso-Sansone, N. V. Prokof'ev, and B. V. Svistunov, Phys. Rev. B {\bf 75}, 134302 (2007).
\bibitem{Sachdev_2011} S. Sachdev, {\it Quantum Phase Transition} (Cambridge University Press, Cambridge, UK, 2011), 2nd ed.
\bibitem{Nakayama_2015} T. Nakayama, I. Danshita, T. Nikuni, and S. Tsuchiya, Phys. Rev. A {\bf 92}, 043610 (2015).
\bibitem{Altman_2015} E. Altman, arXiv:1512.0870 [cond-mat.quant-gas].
\bibitem{Stoferle_2004} T. St\"{o}ferle, H. Moritz, C. Schori, M. K\"{o}hl, and T. Esslinger, Phys. Rev. Lett. {\bf 92}, 130403 (2004).
\bibitem{Endres_thesis} M. Endres, {\it Probing correlated quantum many-body systems at the single-particle level} (Springer, Switzerland, 2014).
\bibitem{Fedichev_1996} P. O. Fedichev, Y. Kagan, G. V. Shlyapnikov, and J. T. M. Walraven, Phys. Rev. Lett. {\bf 77}, 2913 (1996).
\bibitem{Theis_2004} M. Theis, G. Thalhammer, K. Winkler, M. Hellwig, G. Ruff, R. Grimm, and J. H. Denschlag, Phys. Rev. Lett. {\bf 93}, 123001 (2004).
\bibitem{Yamazaki_2010} R. Yamazaki, S. Taie, S. Sugawa, and Y. Takahashi, Phys. Rev. Lett. {\bf 105},  050405 (2010).
\bibitem{Bauer_2009} D. M. Bauer, M. Lettner, C. Vo, G. Rempe, and S. D\"urr, Nat. Phys. {\bf 5}, 339 (2009).
\bibitem{Clark_2015} L. W. Clark, L.-C. Ha, C.-Y. Xu, and C. Chin, Phys. Rev. Lett. {\bf 115}, 155301 (2015).
\bibitem{Holstein_1940} T. Holstein and H. Primakoff, Phys. Rev. {\bf 58}, 1908 (1940).
\bibitem{Chernyshev_2009} A. L. Chernyshev and M. E. Zhitomirsky, Phys. Rev. B {\bf 79}, 144416 (2009).
\bibitem{Kita_2006} T. Kita, J. Phys. Soc. Jpn. {\bf 75}, 044603 (2006).
\bibitem{Yukalov_2006} V. I. Yukalov and H. Kleinert, Phys. Rev. A {\bf 73}, 063612 (2006).
\bibitem{Pethick_2008} C. J. Pethick and H. Smith, {\it Bose--Einstein Condensation in Dilute Gases} (Cambridge University Press, Cambridge, UK, 2008).
\bibitem{Griffin_1996} A. Griffin, Phys. Rev. B {\bf 53}, 9341 (1996).
\bibitem{Shohno_1964} N. Shohno, Prog. Theor. Phys. {\bf 31}, 553 (1964).
\bibitem{Abrikosov_1975} A. A. Abrikosov, L. P. Gorkov, and I. E. Dzyaloshinski, {\it Methods of Quantum Field Theory in Statistical Physics} (Dover Publications, New York, 1975).
\bibitem{Lifshitz_1980} E. M. Lifshitz and L. Pitaevskii, {\it Statistical Physics, Part 2} (Pergamon, Oxford, UK, 1980).
\bibitem{Altland_2010} A. Altland and B. D. Simons, {\it Condensed Matter Field Theory} (Cambridge University Press, Cambridge, UK, 2010), 2nd ed.
\bibitem{Pitaevskii_2003} L. Pitaevskii and S. Stringari, {\it Bose--Einstein Condensation} (Oxford University Press, NewYork, 2003).
\bibitem{Dupuis_2011} N. Dupuis, Phys. Rev. E {\bf 83}, 031120 (2011).
\bibitem{Affleck_1992} I. Affleck and G. F. Wellman, Phys. Rev. B {\bf 46}, 8934 (1992).
\bibitem{Qin_2017} Y. Q. Qin, B. Normand, A. W. Sandvik, and Z. Y. Meng, Phys. Rev. Lett. {\bf 118}, 147207 (2017).
\end{thebibliography}

\end{document}